\documentclass{article}
\usepackage[margin=3cm]{geometry}
\usepackage{graphicx}
\usepackage{hyperref}
\usepackage{float}
\usepackage{booktabs}
\usepackage{caption}
\usepackage{amsmath}
\title{deepmriprep: Voxel-based Morphometry (VBM) Preprocessing via Deep Neural Networks}
\small
\author{Lukas Fisch*$^1$ \and Nils R.~Winter$^1$ \and Janik Goltermann$^1$ \and Carlotta Barkhau$^1$ \and Daniel Emden$^1$ \and Jan Ernsting$^{1,2,3}$ \and Maximilian Konowski$^1$ \and Ramona Leenings$^1$ \and Tiana Borgers$^1$ \and Kira Flinkenflügel$^1$ \and Dominik Grotegerd$^1$ \and Anna Kraus$^1$ \and Elisabeth J. Leehr$^1$ \and Susanne Meinert$^1$ \and Frederike Stein$^4$ \and Lea Teutenberg$^4$ \and Florian Thomas-Odenthal$^4$ \and Paula Usemann$^4$ \and Marco Hermesdorf$^5$ \and Hamidreza Jamalabadi$^4$ \and Andreas Jansen$^4$ \and Igor Nenadic$^4$ \and Benjamin Straube$^4$ \and Tilo Kircher$^4$ \and Klaus Berger$^5$ \and Benjamin Risse$^2$ \and Udo Dannlowski$^1$ \and Tim Hahn$^1$}

\date{\small$^1$University of Münster, Institute for Translational Psychiatry, Münster, Germany \\
$^2$Institute for Geoinformatics, University of Münster, Germany \\
$^3$Faculty of Mathematics and Computer Science, University of Münster, Germany \\
$^4$Department of Psychiatry and Psychotherapy, University of Marburg, Germany \\
$^5$Institute of Epidemiology and Social Medicine, University of Münster, Germany \\
**Correspondence: Lukas Fisch, Albert-Schweitzer-Campus 1, 48149 Münster, Germany. \nolinkurl{l.fisch@uni-muenster.de}\\[2ex]} 
\begin{document}

\maketitle

\begin{abstract}
Voxel-based Morphometry (VBM) has emerged as a powerful approach in neuroimaging research, utilized in over 7,000 studies since the year 2000. Using Magnetic Resonance Imaging (MRI) data, VBM assesses variations in the local density of brain tissue and examines its associations with biological and psychometric variables. Here, we present deepmriprep, a neural network-based pipeline that performs all necessary preprocessing steps for VBM analysis of T$_1$-weighted MR images using deep neural networks. Utilizing the Graphics Processing Unit (GPU), deepmriprep is 37 times faster than CAT12, the leading VBM preprocessing toolbox. The proposed method matches CAT12 in accuracy for tissue segmentation and image registration across more than 100 datasets and shows strong correlations in VBM results. Tissue segmentation maps from deepmriprep have over 95\% agreement with ground truth maps, and its non-linear registration, using supervised SYMNet, predicts smooth deformation fields comparable to CAT12. The high processing speed of deepmriprep enables rapid preprocessing of extensive datasets and thereby fosters the application of VBM analysis to large-scale neuroimaging studies and opens the door to real-time applications. Finally, deepmriprep’s straightforward, modular design enables researchers to easily understand, reuse, and advance the underlying methods, fostering further advancements in neuroimaging research. deepmriprep can be conveniently installed as a Python package and is publicly accessible at \url{https://github.com/wwu-mmll/deepmriprep}.
\end{abstract}


\newpage

\section{Introduction}

Voxel-based Morphometry (VBM) is a widely used analytical approach in neuroimaging research that aims to measure differences in the local concentration of brain tissue across multiple brain images and investigate their association with biological and psychometric variables \cite{ashburner_voxel-based_2000, friston_statistical_1994}. Comparing neuroimaging data is challenging, since the intensity of Magnetic Resonance (MR) images is not standardized, and brain structures differ across individuals. Standard VBM preprocessing addresses this by segmenting the MR images into tissue classes and spatially normalizing the resulting tissue map to a template \cite{spm, ants, afni, freesurfer, cat12, fsl}. Finally, generalized linear models (GLM) are fitted for each voxel, modelling associations between spatially normalized tissue probabilities and the considered biological (e.g. age, sex) or psychometric variables (e.g. symptom severity or cognitive performance scores). If the GLM shows a significant association, the respective voxel could be a region of interest (ROIs) – i.e. a potential neural correlate of the considered variables.

Given that the effect sizes in VBM-based statistical analyses are typically small \cite{nrw_univariate}, MRI studies with thousands of participants are necessary to obtain accurate measurements \cite{marek2022}. To meet this demand, large-scale studies have expanded to include more than 40,000 subjects \cite{bycroft_uk_2018}, resulting in datasets exceeding 100,000 MR images. However, preprocessing these large datasets with existing toolboxes like CAT12 \cite{cat12} can take weeks of even months on standard hardware, delaying scientific progress. Developing a more computationally efficient VBM preprocessing pipeline could alleviate these processing bottlenecks, allowing researches to focus on the conceptual aspects of their studies and accelerating scientific discovery. Therefore, creating a faster VBM preprocessing pipeline is a critical step forward for structural neuroimaging research.

In recent years, deep learning has emerged as a highly effective approach for various tasks in medical image analysis \cite{shen_deep_2017}, delivering state-of-the-art performance in a wide range of image-related applications such as semantic segmentation \cite{unet}. The neuroimaging community followed this trend, such that now neural network-based tools for brain extraction \cite{deepbet, synthstrip, hdbet}, tissue segmentation \cite{kumar_u-segnet_2018, moeskops_automatic_2015}, registration \cite{voxelmorph, hoffmann_anatomy-specific_2023, iglesias_ready--use_2023, mok_fast_2020} and other neuroimaging specific tasks exist \cite{nnunet}.

However, the adoption of neural network-based methods for preprocessing still lags behind classical toolboxes such as CAT12 \cite{cat12}, SPM \cite{spm} or FreeSurfer \cite{freesurfer} due to two main reasons. First, deep learning tools often underperform in a realistic setting where they are applied to MR images from scanner sites unseen during model training \cite{deepbet}. In line with \cite{hoffmann_synthmorph_2022}, this issue can be addressed by increasing the number of scanner sites \cite{martensson_reliability_2020} and extensive use of data augmentation \cite{hoffmann_anatomy-specific_2023, hoffmann_learning_2021, hoffmann_synthmorph_2022, synthstrip}. Second, neural network-based tools are often specialized to one processing step only. Tools like SynthMorph \cite{hoffmann_synthmorph_2022}, SynthStrip \cite{synthstrip}, and EasyReg \cite{iglesias_ready--use_2023} attempt to resolve this by being integrated into the FreeSurfer toolbox, serving as alternatives for parts of its preprocessing pipeline. However, we are not aware of a toolbox which has been developed from the ground up to fully harness the potential of neural networks across all preprocessing steps needed for VBM analysis.

We present deepmriprep, the first preprocessing pipeline for VBM analysis of structural MRI data which is built to fully leverage deep learning. deepmriprep employs neural networks for all three major preprocessing steps: brain extraction, tissue segmentation and spatial registration with a template. Brain extraction is performed by deepbet \cite{deepbet}, the most accurate existing method to remove non-brain voxels in T$_1$-weighted MRIs of healthy adults. To encompass the full VBM preprocessing, we additionally develop neural networks for tissue segmentation and registration in this work. For tissue segmentation, we use a patch-based 3D UNet approach, inspired by \cite{nnunet}, which also exploits neuroanatomical properties such as hemispheric symmetry. Nonlinear image registration is performed using a custom-tailored variant of SYMNet \cite{mok_fast_2020}, which employs a 3D UNet in conjunction with DARTEL shooting \cite{ashburner_fast_2007} to predict smooth and invertible deformation fields.

The neural networks are trained on 685 MR images compiled from 137 OpenNeuro datasets, utilizing silver ground truths and grouped cross-validation to ensure realistic validation performance. The worst predictions are visually inspected to identify potential weaknesses. To investigate the effect of the preprocessing on the VBM-based statistical analyses, deepmriprep and CAT12 are applied to 4,017 subjects from three cohorts. In the subsequent VBM analyses, associations with biological and psychometric variables are investigated and the correlation of the resulting t-maps based on deepmriprep and CAT12 are analyzed. In conclusion, our results indicate that deepmriprep is 37 times faster than CAT12 while achieving comparable accuracy in the individual preprocessing steps and strongly correlated results in the final VBM analysis.

\newpage

\section{Materials and Methods}
\subsection{Datasets}
This study utilizes existing data from 208 studies published on the OpenNeuro platform \cite{openneuro} for training and validation, utilizing cross-validation, and three test datasets: the Münster Neuroimaging Cohort (MNC), the Marburg-Münster Affective Disorders Cohort Study (FOR2107/MACS), and the BiDirect study. Data availability is governed by the respective consortia. No new data was acquired for this study.

\subsubsection{Training and Validation Datasets} \label{sec:traindata}
\begin{figure}[H]
	\centering
	\includegraphics[width=\textwidth]{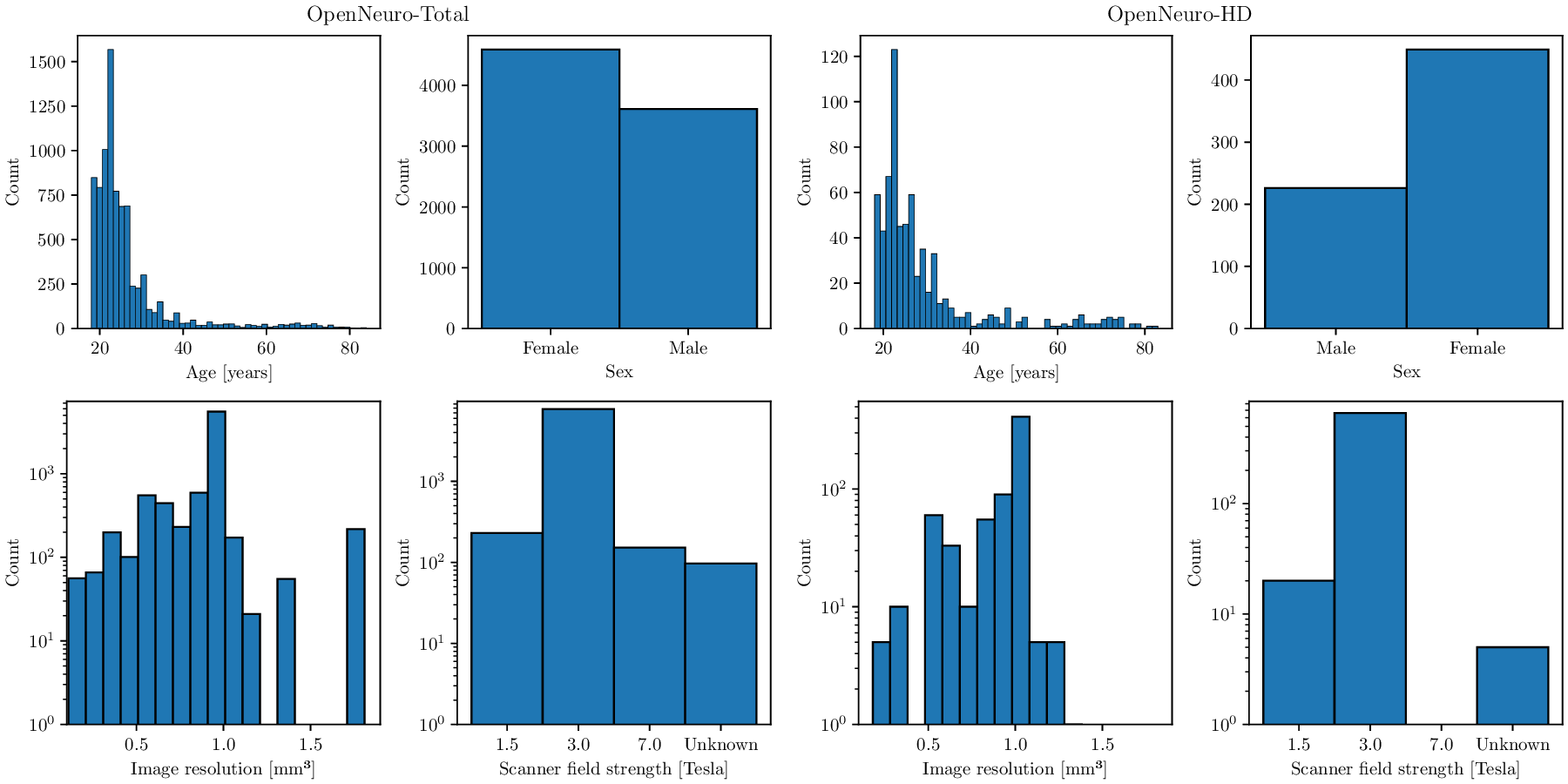}
	\caption{Distribution of age, sex, image resolution, and scanner field strength across 8,279 subjects from OpenNeuro-Total (left) and 685 subjects from OpenNeuro-HD (right).}
	\label{fig:openneuro}
\end{figure}

\paragraph{OpenNeuro-Total}
Out of the over 700 datasets available at OpenNeuro, each dataset that contained at least five T$_1$-weighted (T1w) images from at least five adult healthy controls (HC) was included, resulting in 208 datasets. Based on a successive visual quality check, 29 MR images were excluded, mostly due to improper masking (see Figure \ref{fig:dropouts}A) and erroneous orientation (see Figure \ref{fig:dropouts}B). The remaining compilation of 8,279 T$_1$-weighted MR images is used as the OpenNeuro-Total dataset (see Figure \ref{fig:openneuro}, left and Table \ref{tab:scanner}).

\paragraph{OpenNeuro-HD}
All 8,279 MR images from OpenNeuro-Total were preprocessed using the commonly used CAT12 toolbox (\url{https://neuro-jena.github.io/cat/}) with default parameters \cite{cat12}. To ensure high quality of the silver ground truths, strict quality thresholds were set based on the preprocessing quality ratings provided by the toolbox. To be included, all ratings had to be at least a B- grade, which resulted in the following thresholds: a Surface Euler Number under 25, a Surface Defect Area under 5.0, a Surface Intensity RMSE under 0.1 and a Surface Position RMSE under 1.0. All OpenNeuro datasets that contained fewer than 10 adult HCs after this quality control were excluded. In the remaining datasets, images were ranked according to the Surface Defect Number, and the top five images per dataset that passed a visual quality check were finally included in the dataset. This results in a total of 685 images from 137 datasets called OpenNeuro-HD (see Figure \ref{fig:openneuro}, right and Table \ref{tab:scanner}).

\subsubsection{Test Datasets} \label{sec:testdata}
\begin{figure}[H]
	\centering
	\includegraphics[width=\textwidth]{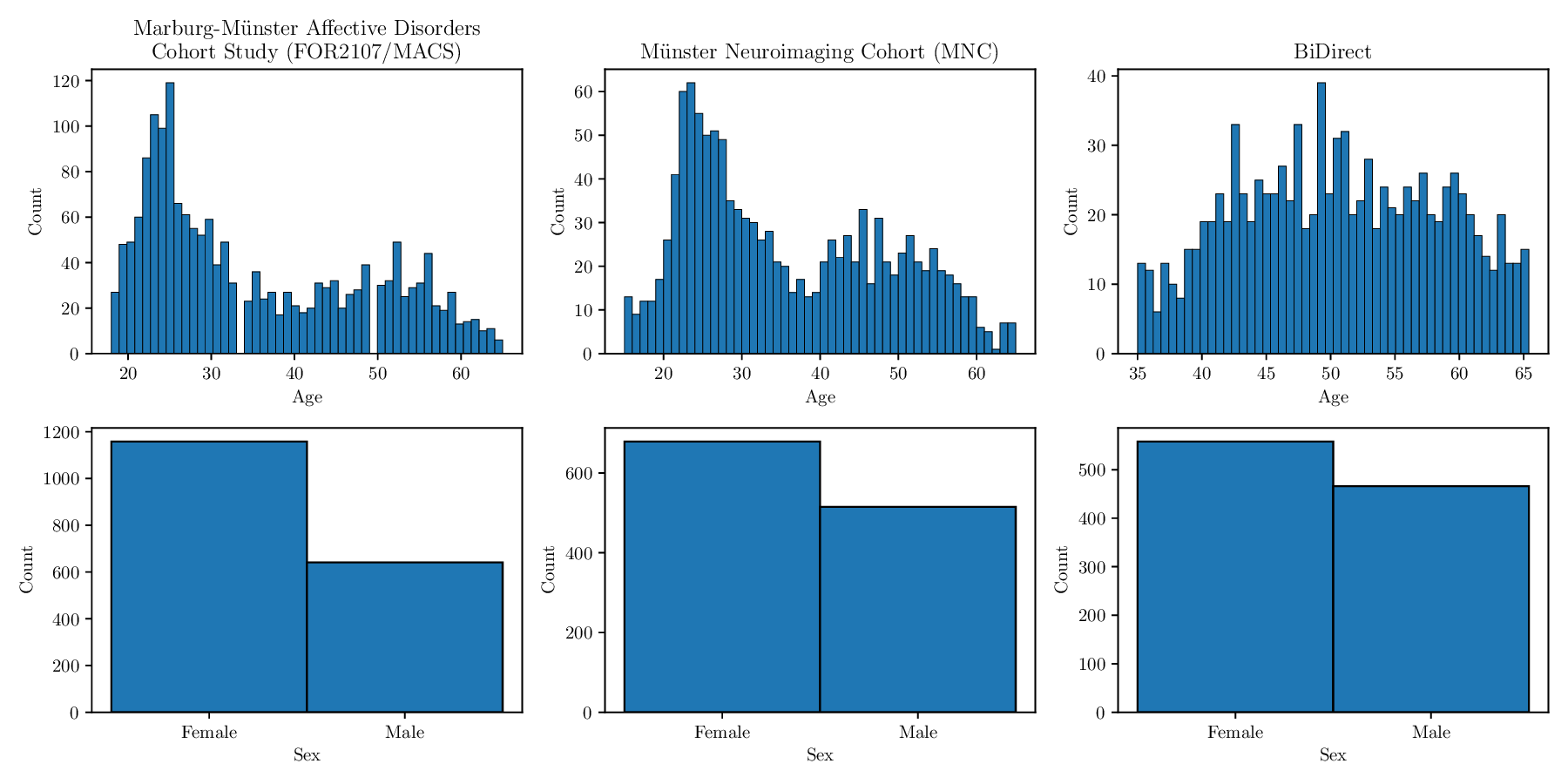}
	\caption{Distribution of age and sex across 1,799 subjects from the Marburg-Münster Affective Disorders Cohort Study (left), 1,194 samples from the Münster Neuroimaging Cohort, and 1,024 subjects from the BiDirect cohort (right).}
	\label{fig:trap}
\end{figure}

For the VBM analyses we use a total of 4,017 MR images from three independent German cohorts (see Figure \ref{fig:trap}) that are not part of the OpenNeuro dataset: the Marburg-Münster Affective Disorders Cohort Study (MACS; N=1,799), the Münster Neuroimaging Cohort (MNC; N=1,194), and the BiDirect cohort (N=1,024). All three cohorts include subsamples with both patients with MDD and HCs who are free from any lifetime mental disorder diagnoses according to DSM-IV criteria.

\paragraph{Marburg-Münster Affective Disorders Cohort Study (FOR2107/MACS)}
Patients were recruited through psychiatric hospitals while the control group was recruited via newspaper advertisements. Patients diagnosed with MDD showed varying levels of symptom severity and underwent various forms of treatment (inpatient, outpatient or none). The FOR2107/MACS was conducted at two scanning sites – University of Münster and University of Marburg. Inclusion criteria for the present study were availability of completed baseline MRI data with sufficient MRI quality. Further details about the structure of the FOR2107/MACS \cite{kircher_neurobiology_2019} and MRI quality assurance protocol \cite{vogelbacher_marburg-munster_2018} are provided elsewhere.

\paragraph{Münster Neuroimaging Cohort (MNC)}
In MNC, patients were recruited from local psychiatric hospitals and underwent inpatient treatment due to a moderate or severe depressive disorder. Further information regarding this study can be found in \cite{dannlowski_disadvantage_2016} and \cite{opel_mediation_2019}.

\paragraph{BiDirect}
The BiDirect Study is a prospective project that comprises three distinct cohorts: patients hospitalized for an acute episode of major depression, patients up to six months after an acute cardiac event, and HCs randomly drawn from the population register of the city of Münster, Germany. Further details on the rationale, design, and recruitment procedures of the BiDirect study have been described in \cite{teismann_establishing_2014} and \cite{teuber_mr_2017}.

\subsection{Input Preprocessing and Ground Truth} \label{sec:preproc}
All datasets are preprocessed using version 12.8.2 of the CAT12 toolbox, which was the latest version available at the time of analysis, with default parameters \cite{cat12}. The affine transformation calculated during this initial CAT12 preprocessing is used such that tissue segmentation (see Section \ref{sec:p0}) and image registration (see Section \ref{sec:warp}) is consistently applied in the templates coordinate space. Image registration is based on gray matter (GM) and white matter (WM) probability maps in the standard resolution of 1.5mm (113x137x113 voxels). 

For tissue segmentation, the unprocessed MR images are affinely registered to the template in a high resolution of 0.5mm (339x411x339 voxels), and the CAT12 preprocessing is repeated based on these high-resolution images. This circumvents any potential image degradation caused by additional resizing of the CAT12 tissue map. Since the MR images are skull-stripped prior to tissue segmentation in deepmriprep’s prediction pipeline (see Section \ref{sec:pred}), all voxels in the MR image which do not contain tissue in the respective tissue map are set to zero. Furthermore, the standard N4 bias correction \cite{n4itk} is applied using the ANTS package \cite{ants} to avoid interference with potential artificial bias fields introduced during data augmentation (see Section \ref{sec:augments}). Finally, min-max scaling between the 0.5th and 99.5th percentile is used as proposed in \cite{nnunet} with one modification: values above the maximum are not clipped to one but scaled via the function $f(x)=1+\log_{10}x$ to prevent any loss of crucial information in areas with extreme intensity values (e.g. blood vessels).

\subsection{Data Augmentation} \label{sec:augments}
Data augmentation is used during training to artificially introduce image artifacts that may occur in real-world datasets. This increases model generalizability, since effects which are infrequent in the training data can be systematically oversampled with any desired intensity. Data augmentations for the image registration step would have to be consist with Equation \ref{eq:diffeo}, requiring specialized implementations. Hence, no data augmentation during the training of the image registration model.

\begin{figure}[H]
	\centering
	\includegraphics[width=\textwidth]{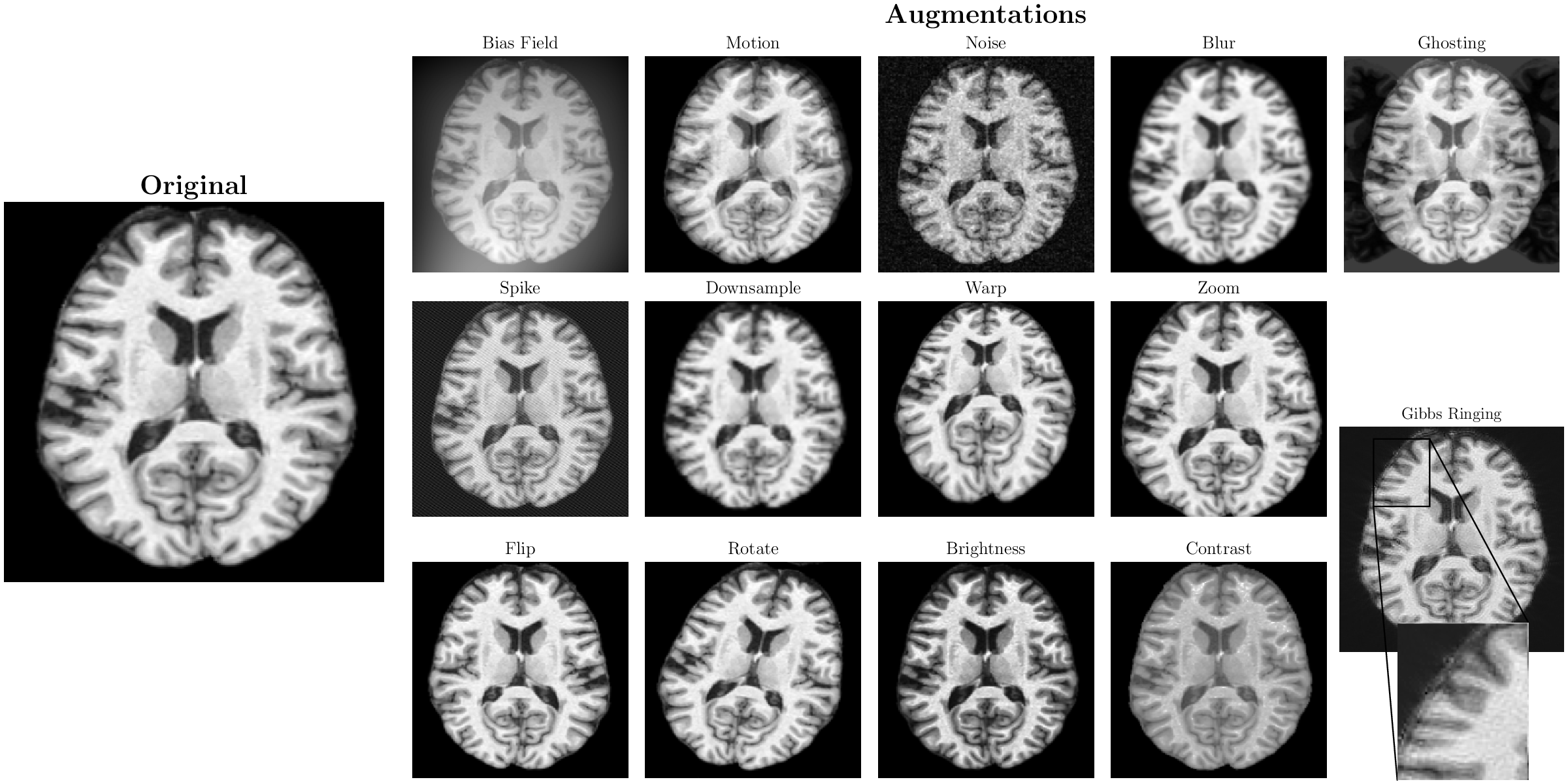}
	\caption{All utilized image augmentations (right) compared to the original image (left).}
	\label{fig:augments}
\end{figure}

The 14 different data augmentations used during model training (see Figure \ref{fig:augments}) can be subdivided into MRI-specific and standard image augmentations. The standard image augmentations consist of four spatial transformations (flip, warp, rotate, and zoom) and two intensity transformations (brightness and contrast). These standard transformations are implemented using the aug\_transforms function of the fastai package \cite{fastai} with max\_rotate set to 15, max\_lighting set to 0.5, max\_warp set to 0.1 and default settings for all remaining parameters.

The MRI-specific augmentations are bias fields, motion artifacts, noise, blurring, ghosting, spike artifacts, downsampling, and Gibbs ringing. Bias fields are simulated with a linear combination of polynomial basis functions \cite{van_leemput_automated_1999} with the order 4, while ghosting \cite{wood_mr_1985}, spike artifacts \cite{graves_body_2013}, and Gibbs ringing \cite{veraart_gibbs_2016} are achieved by introducing artifacts in the k-space of the image. To be MRI-specific, Rician noise \cite{gudbjartsson_rician_1995} instead of standard Gaussian noise is used. The implementation of Rician noise and Gibbs Ringing is based on MONAI \cite{monai}, while all other MRI-specific augmentations are based on the torchio package \cite{torchio}.

\subsection{Tissue Segmentation} \label{sec:p0}
To achieve high-quality tissue segmentation, a cascaded 3D UNet approach, inspired by \cite{nnunet}, is applied to a cropped region of 336x384x336 voxels in the high-resolution MR image (see Section \ref{sec:preproc}). This specific cropping is chosen to make the image dimensions divisible by 16 (required by UNet), without excluding voxels which potentially contain tissue. The first stage of the cascaded UNet processes the whole image with a reduced resolution of 0.75mm (224x256x224 voxels). In the second stage, the original resolution of 0.5mm is processed with a patch-wise approach, which incorporates the prediction from the first stage in its model input. For each patch position, an individual UNet is trained (see Section \ref{sec:p0train}). Both stages of the UNet architecture are identical with respect to the use of the Rectified Linear Unit (ReLU) activation function, instance normalization \cite{ulyanov_instance_2017}, a depth of 4, and the doubling of the number of channels with increasing depth, starting with 8 channels.

\subsubsection{Patchwise UNet} \label{sec:p0patch}
\begin{figure}[H]
	\centering
	\includegraphics[width=\textwidth]{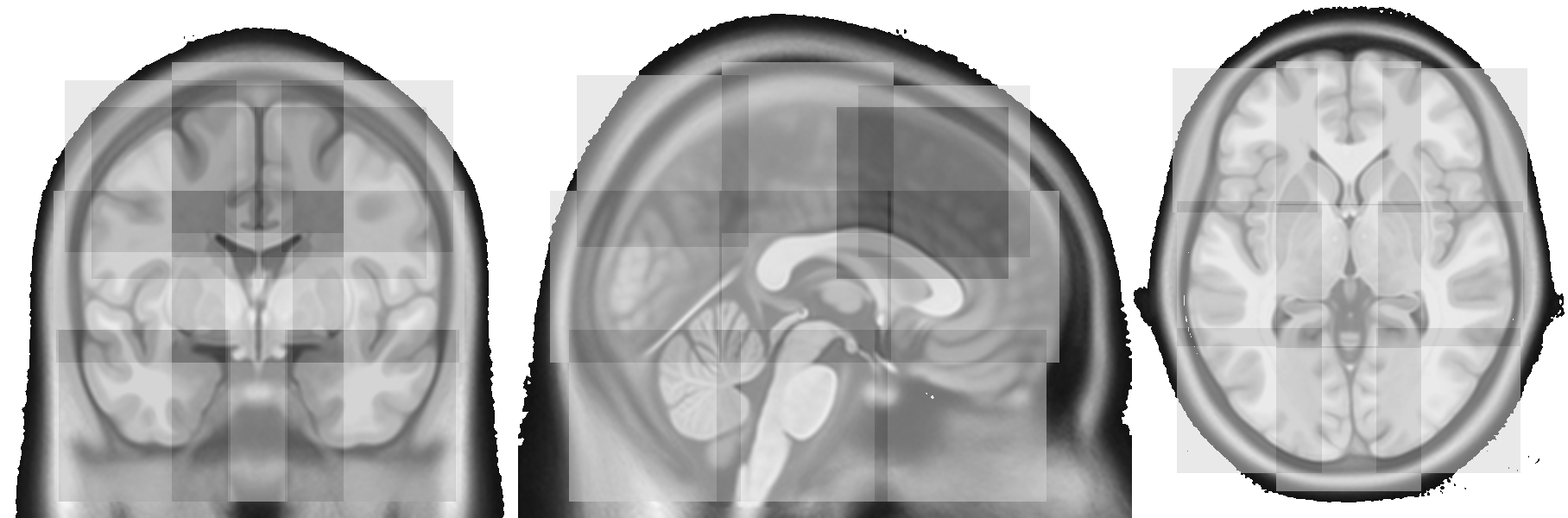}
	\caption{Coronal (left), sagittal (middle), and axial slice (right) of the 128x128x128 voxel patches resulting from the optimized patch positioning. For reference, the T1 template of CAT12, upsampled to the utilized resolution of 0.5mm, is shown in the background.}
	\label{fig:patches}
\end{figure}

The second stage of the cascaded UNet subdivides the 336x384x336 voxels of the high-resolution image into 27 patches, each containing 128x128x128 voxels (see Figure \ref{fig:patches}). For each of these 27 patches, a specific UNet model is trained (see Section \ref{sec:p0train}). To minimize the number of voxels in a patch which do not containing any tissue, the positions of the patches are optimized based on the tissue segmentation masks of all 685 image in OpenNeuro-HD. This iterative optimization starts from a regular grid of 3x3x3 patches which covers the total volume. Then, each patch is moved stepwise by one voxel towards the image centre until this would cause a tissue voxel in one of the 685 images not to be covered by the patch. To exploit the brain's bilateral symmetry, each of the patch on the left hemisphere is moved in lockstep with its corresponding patch on the right hemisphere.

Before applying patches from the right hemisphere to the UNet, we apply flipping along the sagittal axis such that they resemble left hemisphere patches. The resulting prediction is then flipped back. This approach reduces the number of effective patch positions individual UNets need to be trained for from 27 to 18.

Close to the border of a patch, the accuracy of the prediction typically decreases. Therefore, predictions close to the border of a patch are weighted less via Gaussian importance weighting \cite{nnunet} during accumulation of the final prediction containing 336x384x336 voxels.

\subsubsection{Multi-Level Activation Function}
\begin{figure}[H]
	\centering
	\includegraphics[width=\textwidth]{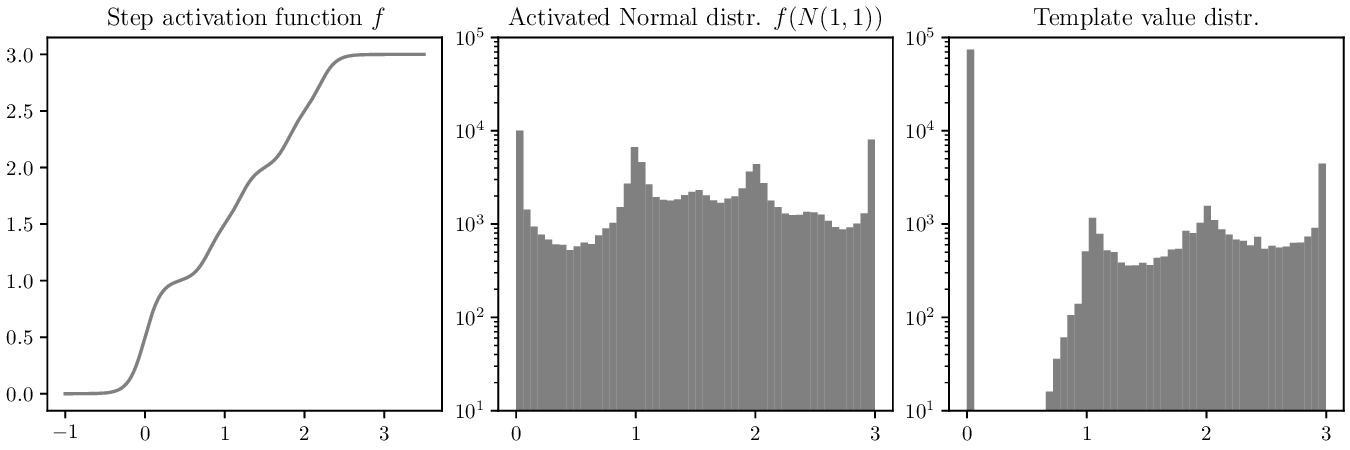}
	\caption{Left: Step activation function $f$ which maps logits to tissue values between 0 and 3. Middle: Activation function $f$ applied to 100,000 values drawn from the normal distribution $N(1,1)$. Right: Tissue value distribution in the MNI template.}
	\label{fig:threestep}
\end{figure}

Based on SPM \cite{spm}, tissue maps outputted by CAT12 contain continuous values between 0 and 3. The values 1, 2 and 3 correspond to cerebrospinal fluid (CSF), GM and WM while 0 indicates that the respective voxel does not contain any tissue. The histogram of the template tissue map (see Figure \ref{fig:threestep}, right) shows that values close to 0, 1, 2, and 3 are more frequent than intermediate values. Furthermore, one can observe smaller peaks around the values 1.5 and 2.5, which correspond to the classes CSF-GM and GM-WM, respectively, which CAT12 introduces. To introduce an inductive bias towards this desired value distribution, the final layer of the tissue segmentation UNet utilizes a custom multi-level activation function inspired by \cite{hu_hierarchical_2019}. This custom multi-level activation is achieved through the summation of six sigmoid functions,
\begin{equation*}
f(x) = S(\alpha x) + \sum_{i \in [1.5, 2., 2.5, 3.]} \frac{S(\alpha \cdot (x-i))}{2} \quad \textrm{with} \quad S(x) = \frac{1}{1+e^{-x}}
\end{equation*}
with $\alpha$ being a parameter of the neural network which is optimized during model training. This function (see Figure 5, left) successfully maps a normal distribution – i.e. typical output distribution of a neural network – to the desired value distribution with peaks at the values 0, 1, 1.5, 2, 2.5, and 3 (see Figure \ref{fig:threestep}, middle). In combination with a mean squared error (MSE) loss, this multi-level activation function facilitates the training of the tissue segmentation model.

\subsubsection{Training Procedure} \label{sec:p0train}
The two stages of 3D UNets are trained in a cascaded fashion. The first stage model is trained for 30 epochs on full-view images with a resolution of 0.75mm (224x256x224 voxels) using a batch size of 1.

The training of the 3x3x3 patch-wise approach in the second stage is more complex and results in 18 models, each dedicated to one of the 18 effective patch positions (3x3x3=27 minus the 9 flipped right hemisphere patches, see Section \ref{sec:p0patch}). Firstly, a UNet is trained on all effective patch positions for one epoch as a foundation model. For each of the 18 patch positions, this foundation model is finally fine-tuned, using solely the respective patch position for 10 epochs. For all patch-based training the batch size is set to 2, and the flip augmentation is disabled for patches located on the left and right hemispheres. The input of the second model consists of patches of the original image, concatenated with the respective patch of the first stage predictions unsampled to the image resolution of 0.5mm. All models are trained with the one-cycle learning rate schedule \cite{onecycle} using a maximal learning rate of 0.001, which follows the default settings of the fastai library \cite{fastai}.

\subsection{Image Registration} \label{sec:warp}
To introduce the neural network-based image registration used for deepmriprep, we first introduce the standard image registration approaches. In standard image registration such as DARTEL \cite{ashburner_fast_2007}, given an input image $\mathbf{I}$ and a template $\mathbf{J}$, the sum of image dissimilarity $D$ and a regularization metric $R$ weighted with a regularization parameter $\Lambda$,
\begin{equation*}
L(\mathbf{I}, \mathbf{J}, \mathbf{\Phi}) = D(\mathbf{I} \cdot \mathbf{\Phi}, \mathbf{J}) + \Lambda R(\mathbf{\Phi})
\end{equation*}
is minimized via the deformation field $\mathbf{\Phi}$. In the loss function $L$ used in this standard approach, the regularization parameter $\Lambda$ controls the trade-off between image similarity and the regularity of the deformation field. In CAT12, the default metric for image similarity is simply the mean squared error (MSE) between the moving and the target image
\begin{equation*}
D(\mathbf{I} \cdot \mathbf{\Phi}, \mathbf{J}) = \mathrm{MSE}(\mathbf{I} \cdot \mathbf{\Phi}, \mathbf{J}) = \frac{1}{| \Omega |} \sum_{\mathbf{p} \in \Omega} | | \mathbf{I} \cdot \mathbf{\Phi} ( \mathbf{p}) - \mathbf{J}(\mathbf{p}) | | ^2
\end{equation*}
and the regularization term is the linear elasticity of the deformation field $\mathbf{\Phi}$

\begin{equation} \label{eq:le}
R(\Phi) = \int \left( \mu \|\epsilon(\Phi)\|^2 + \frac{\lambda}{2} (\text{tr}(\epsilon(\Phi)))^2 \right) \mathrm{d}\mathbf{x}
\end{equation}
where $\mu$ is the weight of the zoom elasticity and $\lambda$ is the weight of the shearing elasticity.

To guarantee that the deformations are invertible, registration frameworks \cite{arsigny_polyrigid_2005, ashburner_fast_2007, camion_geodesic_2001} consider the deformation as the solution of an initial value problem of the form
\begin{equation} \label{eq:diffeo}
\frac{\mathrm{d} \mathbf{\Phi} (s ; \mathbf{x})}{\mathrm{d} s} = \mathbf{v} (\mathbf{\Phi}(s ; \mathbf{x}), s) \quad \textrm{with} \quad \mathbf{\Phi}(0 ; \mathbf{x}) = \mathbf{x}.
\end{equation}
The mapping $\mathbf{x} \rightarrow \mathbf{\Phi}(s ; \mathbf{x})$ defines a family of diffeomorphisms for all time $s \in [0, \tau]$. Hence, it is guaranteed that an inverse of the mapping exists, which can be computed through backwards integration. As proposed in DARTEL, a Stationary Velocity Field (SVF) framework instead of the LDDMM model \cite{beg_computing_2005, vialard_diffeomorphic_2012} allows the velocity field $\mathbf{v}$ to be constant over time. Using this simplification, the regularity of the deformation field - i.e. smoothness and invertibility - is automatically reinforced via forward integration (also called Shooting) of this constant velocity field. This way, a smooth and invertible deformation field can be found by iteratively optimizing the velocity field $\mathbf{v}$ with respect to $L$ using a gradient descent approach.

The SyN registration \cite{avants_symmetric_2008} additionally enforces symmetry between the forwards (image to template) $\mathbf{\Phi}$ and backwards (template to image) deformation field $\mathbf{\Phi}^{-1}$. SyN considers the full forwards and backwards deformations to be compositions of half deformations $\mathbf{\Phi}^{\frac{1}{2}}$ and $\mathbf{\Phi}^{-\frac{1}{2}}$ via
\begin{equation*}
\mathbf{\Phi} = \mathbf{\Phi}^{\frac{1}{2}} \cdot - \mathbf{\Phi}^{-\frac{1}{2}} \quad \textrm{and} \quad \mathbf{\Phi}^ {-1} = \mathbf{\Phi}^{-\frac{1}{2}} \cdot - \mathbf{\Phi}^{\frac{1}{2}}.
\end{equation*}
Based on this consideration, SyN adds the dissimilarity between the image and the backwards deformed template $D(\mathbf{I}, \mathbf{J} \cdot \mathbf{\Phi}^{-1})$ and the dissimilarity between the half forwards deformed image and the half backwards deformed template $D(\mathbf{I} \cdot \mathbf{\Phi}^{\frac{1}{2}}, \mathbf{J} \cdot \mathbf{\Phi}^{-\frac{1}{2}})$ to arrive at a new loss function 
\begin{equation} \label{eq:syn}
L(\mathbf{I}, \mathbf{J}, \Phi) = D(\mathbf{I} \cdot \Phi, \mathbf{J}) + D(\mathbf{I}, \mathbf{J} \cdot \Phi^{-1}) + D(\mathbf{I} \cdot \Phi^{\frac{1}{2}}, \mathbf{J} \cdot \Phi^{\frac{-1}{2}}) + \Lambda R(\Phi).
\end{equation}
Using the diffeomorphic mapping in Equation \ref{eq:diffeo}, velocity fields $\mathbf{v}^{\frac{1}{2}}$ and $\mathbf{v}^{-\frac{1}{2}}$ are used to generate the half deformations $\mathbf{\Phi}^{\frac{1}{2}}$ and $\mathbf{\Phi}^{-\frac{1}{2}}$.

\subsubsection{Model Architecture and Training}
\begin{figure}[H]
	\centering
	\includegraphics[width=\textwidth]{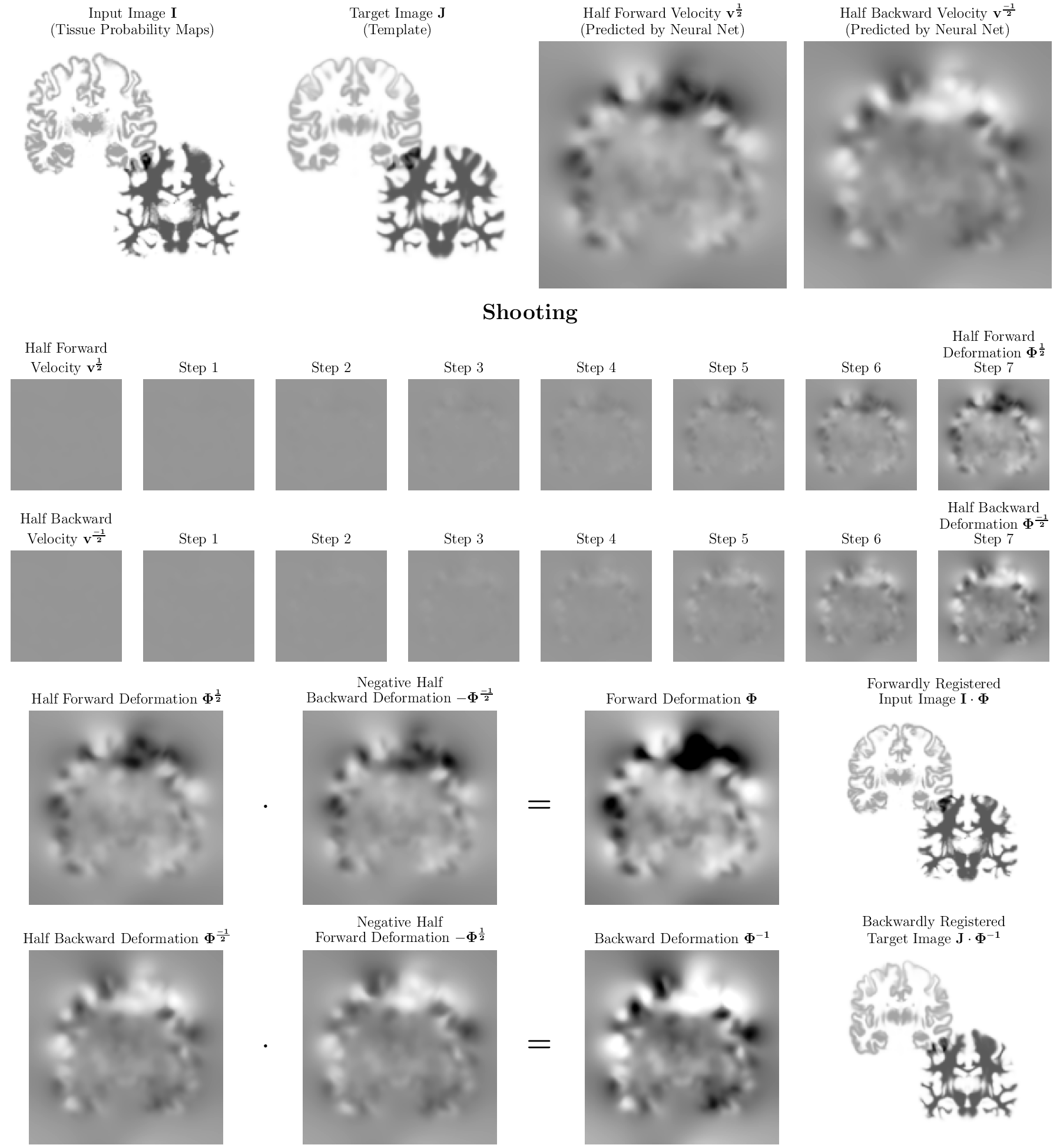}
	\caption{Visualization of diffeomorphic image registration with neural networks. Top: Tissue probability map and registration template (gray matter top left, white matter bottom right) and velocity fields, which are predicted by the neural network. Middle: Stepwise integration (Shooting) of the scaled velocity fields to arrive at the half deformation fields. Bottom: Composition of half deformation fields to form the full forward and backward deformation field. The tissue map is moved into template space (and vice versa) by applying these full deformation fields.}
	\label{fig:syn}
\end{figure}

The neural network-based image registration framework used for deepmriprep is based on SYMNet \cite{mok_fast_2020} and uses a UNet to predict the forward and backward velocity field $\mathbf{v}^{\frac{1}{2}}$ and $\mathbf{v}^{-\frac{1}{2}}$. Analogous to the SyN registration, these velocity fields are integrated according to Equation \ref{eq:diffeo} to arrive at the half deformation fields $\mathbf{\Phi}^{\frac{1}{2}}$ and $\mathbf{\Phi}^{-\frac{1}{2}}$ via the scaling and squaring method \cite{arsigny_polyrigid_2005, ashburner_fast_2007} with $\tau=7$ time steps (see Figure \ref{fig:syn}).

Similar to the neural network architecture used for tissue segmentation (see Section \ref{sec:p0}), the UNet uses instance normalization \cite{ulyanov_instance_2017}, a depth of 4, and the doubling of the number of channels with increasing depth starting with 8 channels. However, we apply two modifications 1. Usage of LeakyReLU \cite{xu_empirical_2015} instead of ReLU activation layers, and 2. Hyperbolic tangent (tanh) activation function in the final layer, ensuring that the UNet’s output conforms to the value range of -1 to 1 used for image coordinates by PyTorch \cite{pytorch}. The model is trained for 50 epochs using the one-cycle learning rate schedule with a maximal learning rate of 0.001.

During initial tests, training with the standard SyN loss function (see Equation \ref{eq:syn}) lead to major artifacts in the predicted velocity and deformation field (see Figure \ref{fig:ablations}B). To avoid these artifacts, we tested supervised approaches which utilize deformation fields created by CAT12 (see Figure \ref{fig:ablations}C-E). Using an iterative approach, we determined the velocity fields $\mathbf{v}_{\textrm{CAT}}^{\frac{1}{2}}$ and $\mathbf{v}_{\textrm{CAT}}^{-\frac{1}{2}}$ which produce these given deformation fields $\mathbf{\Phi}_{\textrm{CAT}}$ and $\mathbf{\Phi}_{\textrm{CAT}}^{-1}$ in our PyTorch-based implementation and used these velocity fields as targets. Using the MSE, the resulting loss function $L_v$ measures disagreements between the predicted velocity fields $\mathbf{v}^{\frac{1}{2}}$ and $\mathbf{v}^{-\frac{1}{2}}$ and the targets via
\begin{equation*}
L_{\textbf{v}}(\mathbf{v}^{\frac{1}{2}}, \mathbf{v}^{-\frac{1}{2}}) = \frac{1}{| \Omega |} \sum_{\mathbf{p} \in \Omega} | | \mathbf{v}_{\textrm{CAT}}^{\frac{1}{2}} ( \mathbf{p}) - \mathbf{v}^{\frac{1}{2}} ( \mathbf{p}) | | ^2 + | | \mathbf{v}_{\textrm{CAT}}^{-\frac{1}{2}} ( \mathbf{p}) - \mathbf{v}^{-\frac{1}{2}} ( \mathbf{p}) | | ^2.
\end{equation*}

Using this loss function, the predicted velocity fields show fewer artifacts, but based on the resulting Jacobi determinant field $\mathbf{J_\Phi}$, some inaccuracies remain (see Figure \ref{fig:ablations}C). The Jacobi determinant indicates the volume change caused by the deformation for each voxel. By explicitly adding the MSE between the predicted and ground truth Jacobi determinant $\mathbf{J_\Phi}$, the loss function
\begin{equation*}
L_{\textbf{v}, \textbf{J}_\Phi}(\mathbf{v}^{\frac{1}{2}}, \mathbf{v}^{-\frac{1}{2}}) = \frac{1}{| \Omega |} \sum_{\mathbf{p} \in \Omega} | | \mathbf{v}_{\textrm{CAT}}^{\frac{1}{2}} ( \mathbf{p}) - \mathbf{v}^{\frac{1}{2}} ( \mathbf{p}) | | ^2 + | | \mathbf{v}_{\textrm{CAT}}^{-\frac{1}{2}} ( \mathbf{p}) - \mathbf{v}^{-\frac{1}{2}} ( \mathbf{p}) | | ^2 + | | \mathbf{J_\Phi}_{\textrm{CAT}} ( \mathbf{p}) - \mathbf{J_\Phi} ( \mathbf{p}) | | ^2
\end{equation*}
improves regularity of the predicted velocity fields and the resulting Jacobi determinant field (see Figure \ref{fig:ablations}D). Finally, we reintroduce the original loss function $L_{\textrm{SyN}}$ as
\begin{equation*}
L_{supervised}(\mathbf{v}^{\frac{1}{2}}, \mathbf{v}^{-\frac{1}{2}}) = L_{\textbf{v}, \textbf{J}_\Phi}(\mathbf{v}^{\frac{1}{2}}, \mathbf{v}^{-\frac{1}{2}}) + \beta L_{\textrm{SyN}}
\end{equation*}
with $\beta$ set to $2 \cdot 10^{-5}$. The fields predicted with this approach, called supervised SYMNet or sSYMNet, do not show any apparent artifacts (see Figure \ref{fig:ablations}E).

\subsection{Evaluation}
We follow best practices by applying a 5-fold cross-dataset validation – i.e. cross-validation with datasets grouped together – using 137 datasets from OpenNeuro-HD (see Section \ref{sec:traindata}). Thereby, we enforce realistic performance measures, since all reported results are achieved in datasets unseen during training of the respective model. We apply the same folds across tissue segmentation, image registration, and GM masking (see Section \ref{sec:nogm}) to avoid data leakage between these processing steps. Test datasets (see Section \ref{sec:testdata}) or datasets of OpenNeuro-Total, which are not included in OpenNeuro-HD (see Section \ref{sec:p0total}), are evaluated with an additional model which is trained with data from all 685 subjects of the OpenNeuro-HD compilation.

Given that the distribution of performance metrics across images is often skewed, the median is used as a measure of central tendency, complemented by a visual inspection of negative outliers.

To evaluate tissue segmentation and GM masking performance, we use the Dice score $DSC$. The image registrations are evaluated based on the regularity of the deformation field and the dissimilarity between the warped input and template image. This dissimilarity is measured using the voxel-wise mean squared error $MSE$ between the images. The deformation field’s regularity - i.e. its physical legitimacy - is quantified via the linear elasticity $LE$ (see Equation \ref{eq:le}).

\subsection{Prediction Pipeline} \label{sec:pred}

The full deepmriprep pipeline used prior to the GLM analysis in Section \ref{sec:vbm} consists of six steps: Brain extraction, affine registration, tissue segmentation, tissue separation, nonlinear registration, and smoothing. After brain extraction via deepbet \cite{deepbet} with default settings, affine registration is applied using the sum of the mean squared error (between image and template) and Dice Loss (between image brain mask and template brain mask). The affine registration is implemented in torchreg \cite{torchreg} with zero padding – sensible after brain extraction – and the default two-staged setting with 500 iterations in 12mm³ and successive 100 iterations in 6mm³ image resolution. After tissue segmentation (see Section \ref{sec:p0}) and prior to image registration (see Section \ref{sec:warp}), we apply GM masking in the ventricles and around the brain stem to conform the probability masks to an undocumented step in the existing CAT12 preprocessing (see Section \ref{sec:nogm}). After image registration of the GM and WM probability masks, Gaussian smoothing with a 6mm full width at half maximum (FWHM) kernel is applied. In line with all prior steps, smoothing (a simple convolution operation) is implemented in PyTorch, enabling GPU acceleration throughout the entire prediction pipeline.

\subsection{VBM Analyses}
To investigate the effect of the preprocessing on the VBM analyses, deepmriprep- and CAT12-preprocessed data is used to examine statistical associations with both biological - i.e. age, sex and Body mass index (BMI) - and psychometric variables - i.e. Years of Education, MDD vs HC and intelligence quotient (IQ). To ensure the reliability of results, each VBM analysis is repeated 100 times with a randomly picked 80\% subset of the total test dataset compilation. Finally, the median t-map across these 100 VBM analyses is used to compare the VBM results of deepmriprep and CAT12.

\section{Results}
\subsection{Tissue Segmentation}
\subsubsection{OpenNeuro-HD}
\begin{figure}[H]
	\centering
	\includegraphics[width=\textwidth]{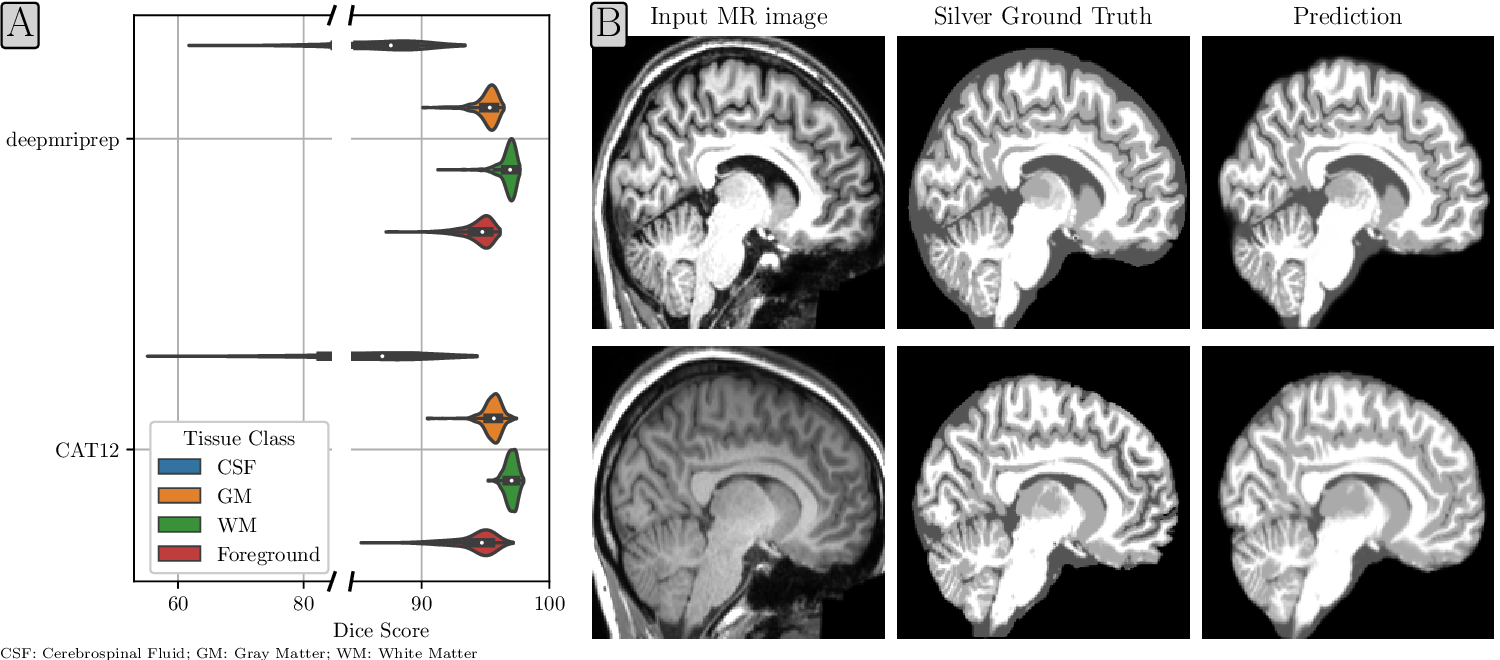}
	\caption{Dice scores of deepmriprep and CAT12 with respect to the cerebrospinal fluid (CSF), gray matter (GM), white matter (WM) and foreground - i.e. mean of CSF, GM and WM. Right: Sagittal slice of the T$_1$-weighted MR image, the silver ground truth and the predicted tissue segmentation map in the sample which resulted in the lowest foreground Dice score for deepmriprep (first row) and CAT12 (second row).}
	\label{fig:p0}
\end{figure}

During cross-dataset validation, deepmriprep demonstrated robust tissue segmentation, achieving a median Dice score $DSC_{\textrm{median}}$ of 94.8 across validation images (see Figure \ref{fig:p0}A and Table \ref{tab:results} left). This high level of agreement with the ground truth - i.e. CAT12 tissue segmentation maps based on high resolution images (see Section \ref{sec:preproc}) - was comparable to CAT12 (in original image resolution), which achieves a median Dice score $DSC_{\textrm{median}}$ of 94.7. GM and WM tissue was segmented with high accuracy with both deepmriprep ($DSC_{\textrm{median}}^{\textrm{GM}}=95.3$, $DSC_{\textrm{median}}^{\textrm{WM}}=96.9$) and CAT12 ($DSC_{\textrm{median}}^{\textrm{GM}}=95.7$, $DSC_{\textrm{median}}^{\textrm{WM}}=97.0$). Segmentation of CSF resulted in the lowest median Dice scores $DSC_{\textrm{median}}^{\textrm{CSF}}$ of 87.6 for deepmriprep and 86.9 for CAT12. Additionally, the Dice scores of CSF showed the strongest outliers across all 685 validation images, with minimal Dice scores $DSC_{\textrm{min}}^{\textrm{CSF}}$ of 61.8 for deepmriprep and 55.1 for CAT12. In the tissue maps resulting in the minimal foreground Dice score $DSC_{\textrm{min}}$ for each method (see Figure \ref{fig:p0}B), deepmriprep and CAT12 produced a thinner outer layer of CSF than the ground truth. With respect to GM and WM, the tissue maps of both methods showed no notable visual differences to the ground truth.

\begin{table}[H]
\caption{Median Dice score ($\pm$ standard deviation) of deepmriprep and CAT12 with respect to the cerebrospinal fluid (CSF), gray matter (GM), white matter (WM) and foreground - i.e. mean of CSF, GM and WM - in the OpenNeuro-HD (left) and OpenNeuro-Total dataset (right).}
\label{tab:results}
\centering
\begin{tabular}{l|cc}
Tissue Class & \textbf{deepmriprep} & \textbf{CAT12} \\
\midrule
CSF & 87.6 $\pm$ 4.3 & 86.9 $\pm$ 6.1 \\
GM & 95.3 $\pm$ 0.9 & 95.7 $\pm$ 0.7 \\
WM & 96.9 $\pm$ 0.7 & 97.0 $\pm$ 0.4 \\
Foreground & 94.8 $\pm$ 1.2 & 94.7 $\pm$ 1.7
\end{tabular}
\quad
\begin{tabular}{l|c}
Tissue Class & \textbf{deepmriprep vs. CAT12} \\
\midrule
CSF & 84.0 $\pm$ 6.9 \\
GM & 94.1 $\pm$ 3.7 \\
WM & 96.4 $\pm$ 3.4 \\
Foreground & 91.5 $\pm$ 4.1 \\
GM-WM-Mean & 95.3 $\pm$ 3.4
\end{tabular}
\end{table}

\subsubsection{OpenNeuro-Total} \label{sec:p0total}

\begin{figure}[H]
	\centering
	\includegraphics[width=\textwidth]{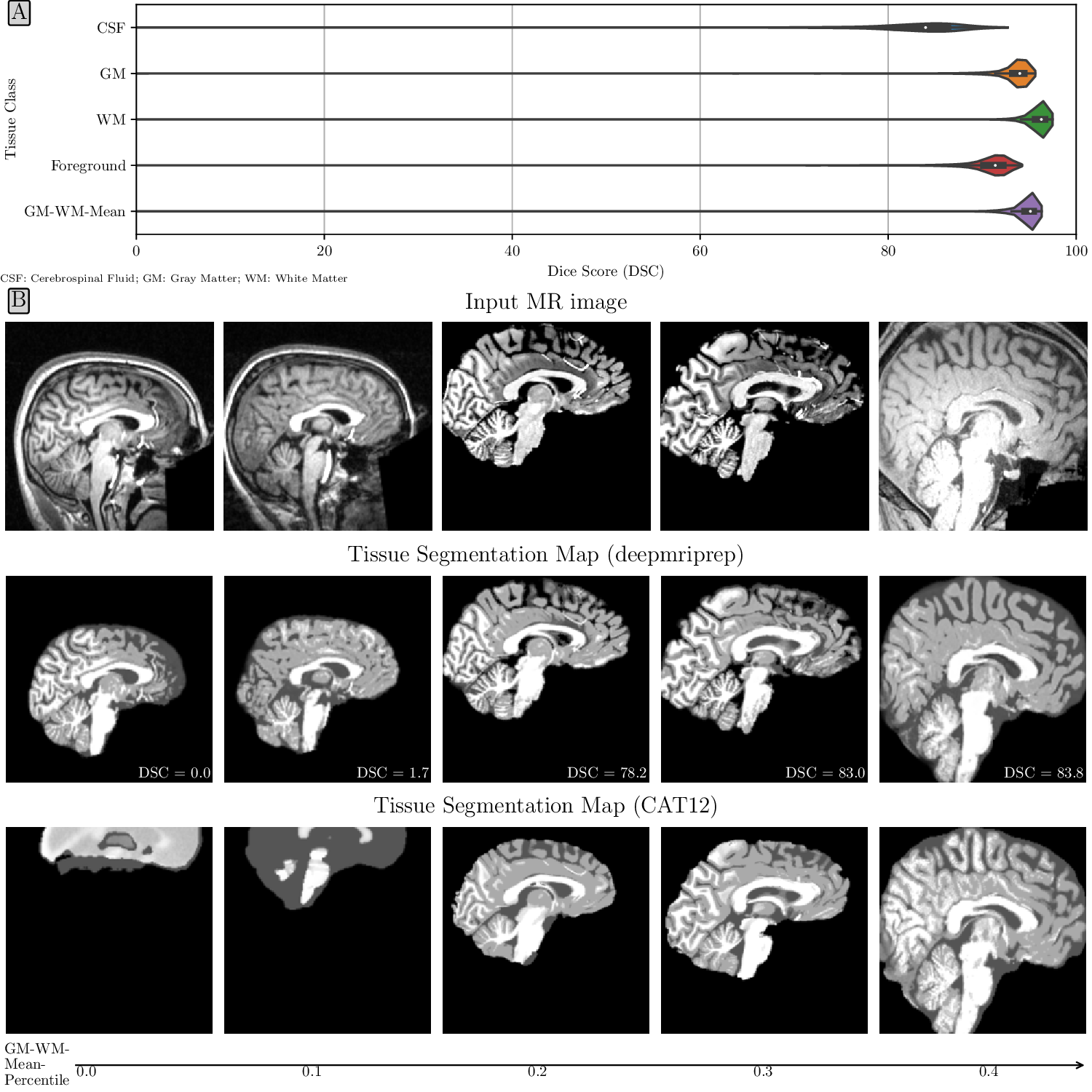}
	\caption{Top: Dice scores of deepmriprep across all 8,279 images from OpenNeuro and CAT12 with respect to the cerebrospinal fluid (CSF), gray matter (GM), white matter (WM), foreground - i.e. mean of CSF, GM and WM – and the mean of GM and WM (GM-WM-Mean). Bottom: MR image input (first row) and tissue segmentation map (deepmriprep: second row, CAT12: thirs row) which resulted in the 0.0, 0.1, 0.2, 0.3 and 0.4 percentile GM-WM Dice scores across all 8,279 images.}
	\label{fig:p0total}
\end{figure}

The OpenNeuro-Total compilation contains 8,279 MR images from 208 datasets which, in contrast to OpenNeuro-HD, did only undergo a minimal quality control. Despite this challenging setting, the median Dice score $DSC_{\textrm{median}}$ of 91.5 between deepmriprep and CAT12 tissue maps – not to be confused with Dice scores of each tool’s output compared to ground truths – showed high agreement for most of the respective tissue maps (see Figure \ref{fig:p0total}A). Again, GM and WM segmentation was most consistent with Dice scores of $DSC_{\textrm{median}}^{\textrm{GM}}=94.1$ and $DSC_{\textrm{median}}^{\textrm{WM}}=96.4$ while CSF segmentation resulted in a lower median Dice score of $DSC_{\textrm{median}}^{\textrm{CSF}}=84.0$.

Despite the absence of ground truths, visually comparisons of tissue maps with low Dice scores – i.e. low agreement between deepmriprep and CAT12 – enables qualitative assessment of each tools’ robustness. Since most VBM analyses are not concerned with CSF, we selected the tissue maps based on the mean Dice score of GM and WM $DSC^{\textrm{GWM}}$. 

Throughout the tissue maps with the 0.0th, 0.1th, 0.2th, 0.3th and 0.4th percentile of GM and WM Dice score $DSC^{\textrm{GWM}}$ (see Figure \ref{fig:p0total}B), deepmriprep showed reasonable results with minor artifacts, while CAT12 was prone to errors. In the 0.0th and 0.1th percentile tissue maps, CAT12 produced unusable results with respective Dice scores of $DSC^{\textrm{GWM}}=0.0$ and $DSC^{\textrm{GWM}}=1.7$, compared to the reasonable tissue maps created by deepmriprep. In the 0.2th and 0.3th percentile tissue maps, CAT12 produced less detailed tissue maps than deepmriprep and misclassified tissue at the edge of the brain as background. deepmriprep properly classified the outer edge tissue, except for an area of CSF which was misclassified as background. In the 0.4th percentile tissue map, the strong bias field present in the MR image resulted in tissue maps with reduced quality for both tools. The same characteristic sources of errors could be found across the 16 tissue maps with the lowest agreement between deepmriprep and CAT12 (see Figure \ref{fig:p0totalsupp}), again measured by $DSC^{\textrm{GWM}}$. Finally, due to an error, CAT12 did not produce any tissue map for one MR image, while deepmriprep processed all 8,279 images without any errors.

\subsection{Image Registration}
\begin{figure}[H]
	\centering
	\includegraphics[width=\textwidth]{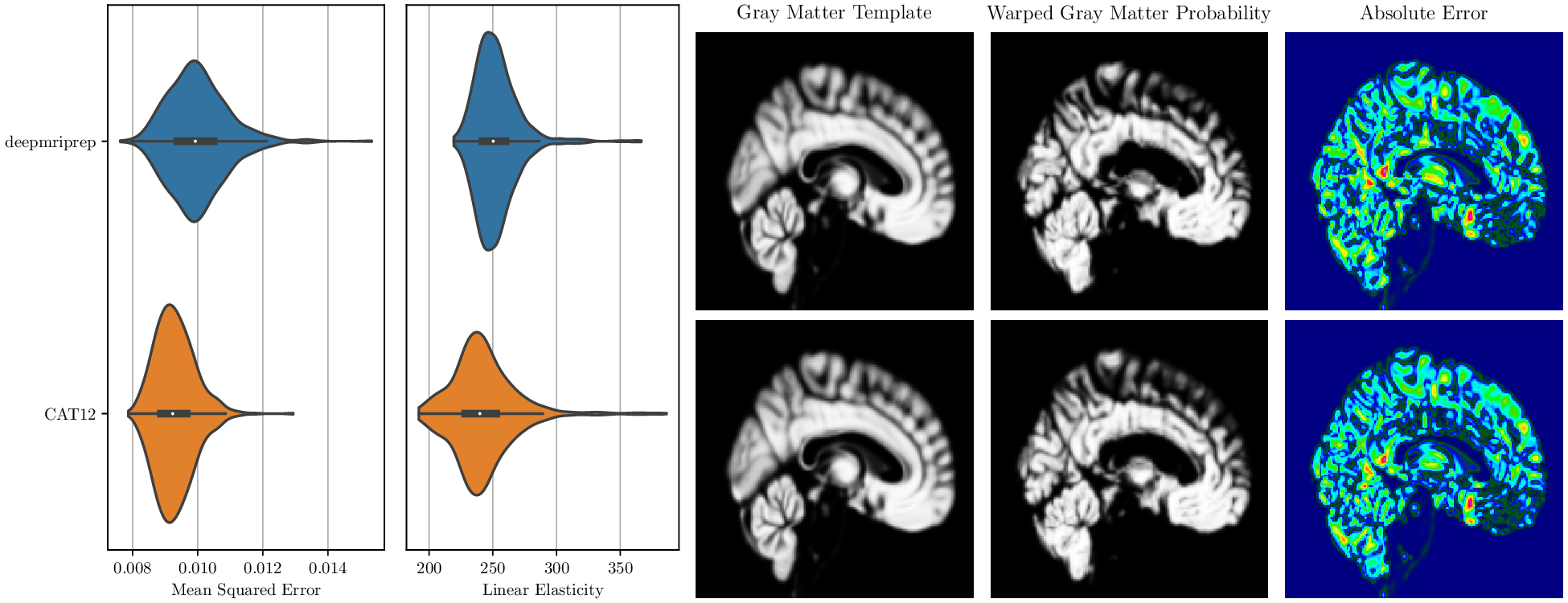}
	\caption{Left: Image registration metrics Mean Squared Error and Linear Elasticity of deepmriprep and CAT12 across all 685 images in OpenNeuro-HD. Right: Gray matter template, next to warped output and the resulting absolute error between template and output of deepmriprep (top) and CAT12 (bottom).}
	\label{fig:warp}
\end{figure}

The registration of tissue probability maps with deepmriprep resulted in a median mean squared error $MSE_{\textrm{median}}$ of $9.9 \cdot 10^{-3}$ and a median Linear Elasticity $LE_{\textrm{median}}$ of 250 during cross-dataset validation (see Figure \ref{fig:warp}, left). These metrics indicate deepmriprep being on par with CAT12 ($MSE_{\textrm{median}} = 9.2 \cdot 10^{-3}$, $LE_{\textrm{median}}=240$). While CAT12 showed slightly better median metrics, the supervised SYMNet used within deepmriprep resulted in a smaller maximal Linear Elasticity across images $LE_{\textrm{max}}$ of 366 (CAT12: $LE_{\textrm{max}}=386$). This favourable Linear Elasticity indicates a better regularity of the deformation field for challenging probability maps – i.e. maps which show large voxel-wise differences to the template.

For both registration methods, the same probability map resulted in the largest voxel-wise MSE after registration (see Figure \ref{fig:warp}, right). Visual inspection of this warped probability map uncovers a small misalignment at the upper edge of the ventricles for deepmriprep, indicating less rigour in aligning the map with the template. Based on the absolute voxel-wise difference to the template, no apparent differences between deepmriprep and CAT12 could be found.

\subsection{VBM Analyses} \label{sec:vbm}
\vspace{-8pt}
\begin{figure}[H]
	\centering
	\includegraphics[width=0.8\textwidth]{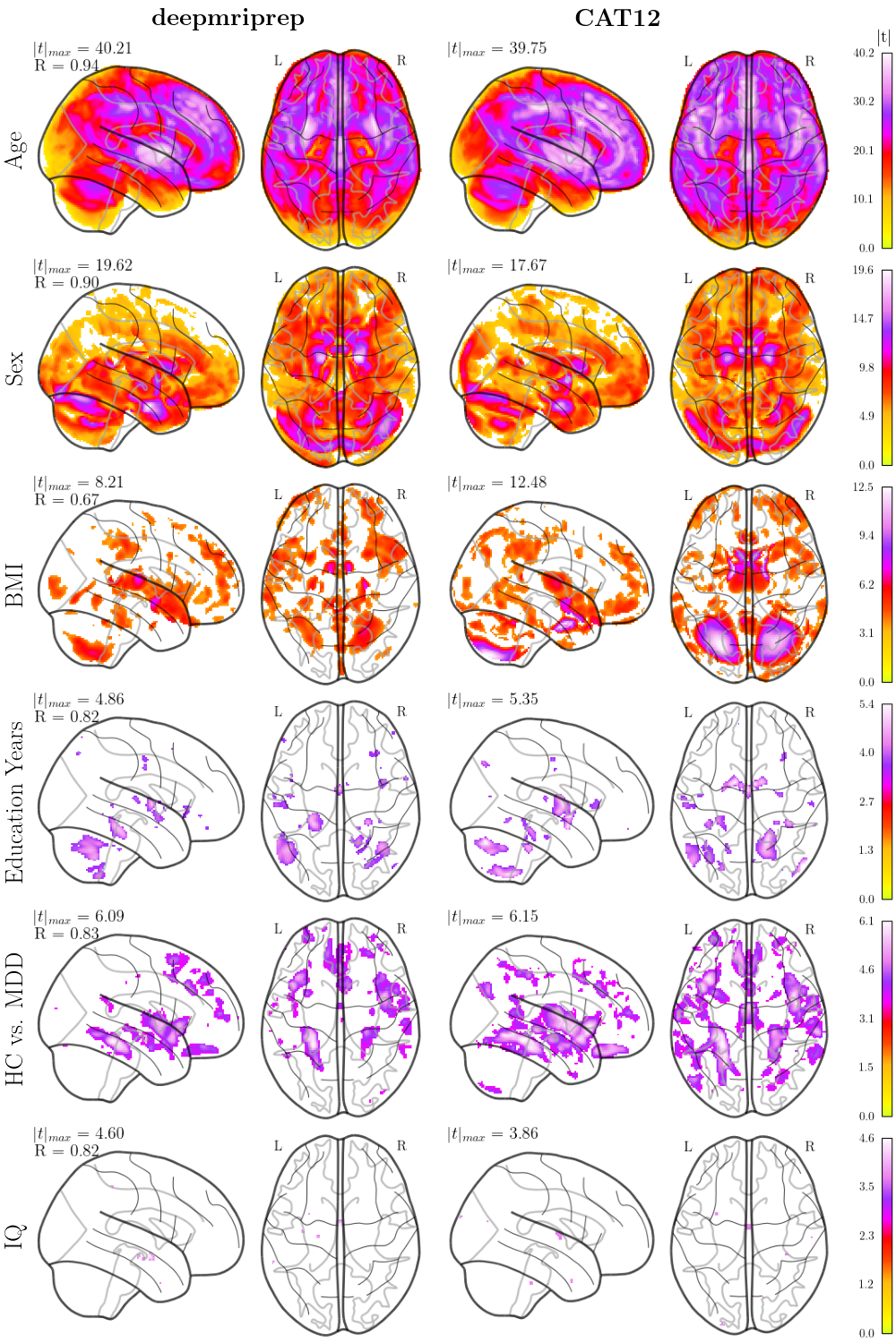}
	\caption{Absolute t-scores of GLM analysis between gray matter volume and age, sex, Body mass index (BMI), years of education, HC vs. MDD (healthy control vs. major depressive disorder) and intelligence quotient (IQ) based on deepmriprep- (left) and CAT12-preprocessing (right) thresholded at $p<0.001$. The respective maximum values and correlation coefficients between deepmriprep and CAT12 are based on unthresholded absolute t-scores.}
	\label{fig:tscores}
\end{figure}

VBM analysis results in gray matter based on deepmriprep and CAT12 demonstrated high similarity (see Figure \ref{fig:tscores}), with strong correlation between the respective t-maps. The correlation of t-maps remained strong even for the psychometric variables - years of education, HC vs. MDD and IQ – despite their smaller effects compared to biological variables – i.e. age, sex and BMI. All analyses resulted in correlation coefficients of $r > 0.8$ with BMI ($r=0.67$) being the only exception. The equivalence between deepmriprep- and CAT12-based analysis outcomes is also supported by their similar maximal, absolute t-scores $|t|_{\textrm{max}}$, especially for age and HC vs. MDD. deepmriprep resulted in a larger maximal, absolute t-score for sex and IQ and smaller maximal, absolute t-scores for years of education and BMI. The difference in maximal values and the reduced t-map correlation for BMI was primarily caused by a large cluster in the outer cerebellum, which was only found based on the CAT12 preprocessing (see Figure \ref{fig:tscores}). This cluster could also be found in the VBM results in white matter (see Figure \ref{fig:tscoreswm}) which otherwise also exhibit strong correlations between t-maps of $r > 0.8$. 

\begin{figure}[H]
	\centering
	\includegraphics[width=\textwidth]{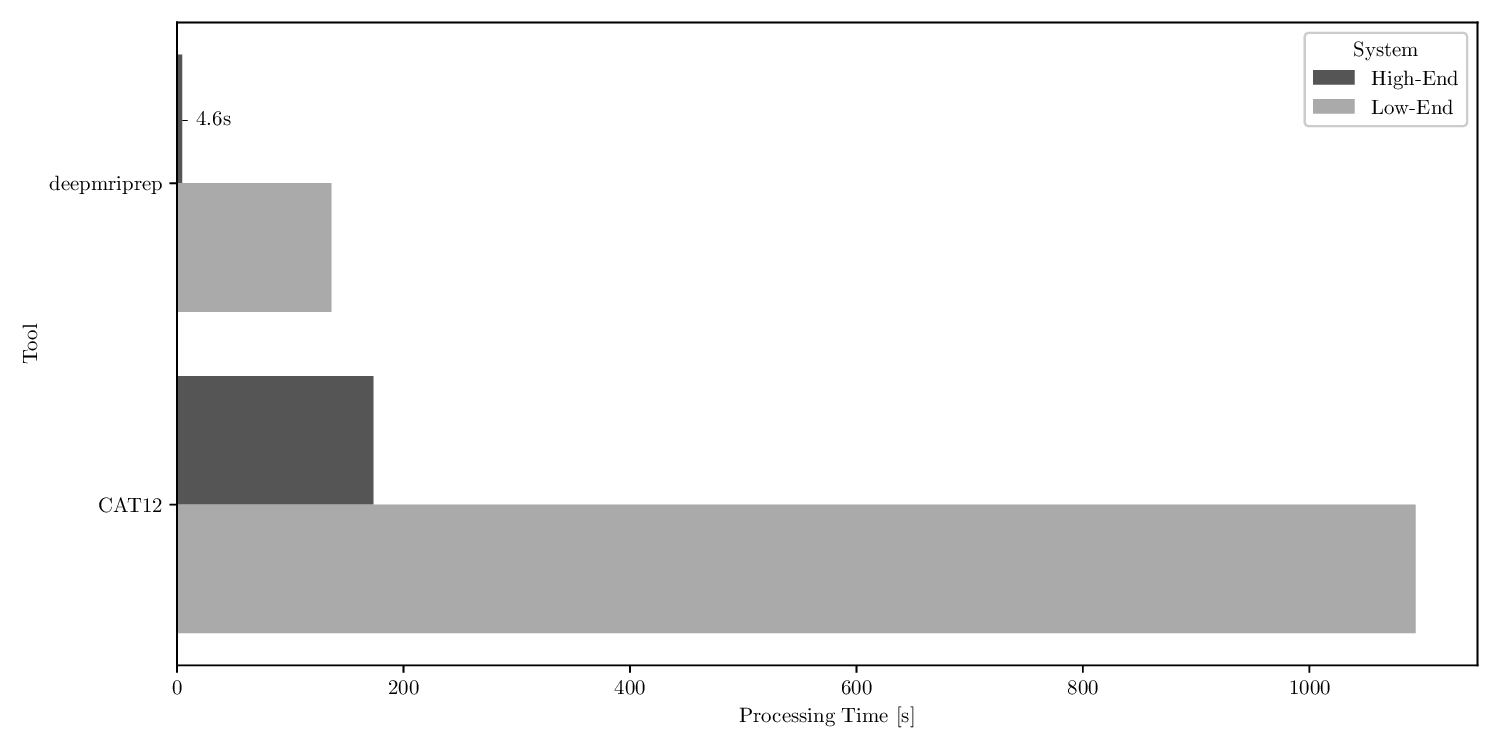}
	\caption{Processing time in seconds $s$ taken per MR image of deepmriprep and CAT12. The high-end system utilizes a AMD Ryzen 9 5950X, a NVIDIA GeForce RTX 3090 Graphics Processing Unit (GPU) and 128GB of memory while the low-end system uses a Intel i7-8565U without a dedicated GPU and 16GB of memory.}
	\label{fig:speed}
\end{figure}

\subsection{Processing Time}

deepmriprep achieved the highest processing speed on both low-end and high-end hardware (see Figure \ref{fig:speed}). On high-end hardware deepmriprep took averagely 4.6 seconds per MR image utilizing the Graphics Processing Unit (GPU), while CAT12 parallelized over all 16 cores of the high-end processor took 173 seconds on average per MR image. On low-end hardware, deepmriprep and CAT12 take 136 second and 1096 seconds per MR image, respectively.

\section{Discussion}
We present deepmriprep, a neural network-based pipeline designed to perform all preprocessing steps necessary for VBM analysis of T$_1$-weighted MR images. By leveraging efficient neural networks in combination with Graphics Processing Unit (GPU) support throughout the entire pipeline, deepmriprep achieves processing speeds 37 times faster than CAT12, a leading toolbox known to be the more efficient preprocessing alternative to FreeSurfer. The results indicate on par accuracy with CAT12 regarding tissue segmentation and image registration across 137 datasets unseen during respective model training. Comparisons of VBM analyses based on deepmriprep and CAT12 reveal strong correlation in the resulting t-maps, even for small effect sizes across various psycholometric and biological variables.

Gray matter and white matter segmentation maps produced by deepmriprep exhibit over 95\% agreement with high-quality ground truth maps from the OpenNeuro-HD dataset compilation, with median Dice scores of 95.3 and 96.9 for gray and white matter, respectively. Visual inspection of tissue maps from the OpenNeuro-Total dataset compilation shows that deepmriprep delivers reliable segmentation even for MR images that yield incorrect or no output from CAT12. Across the 8,279 MR images from the 208 datasets in OpenNeuro-Total, the few remaining inaccuracies are typically the exclusion of CSF in the brain and omission of fine details in the white matter of the cerebellum.

For nonlinear registration of tissue probability maps, deepmriprep builds upon the existing SYMNet architecture which is designed to predict smooth and physically plausible deformation fields. Despite its design, the baseline SYMNet produced deformation fields with apparent irregularities. The supervised learning approach in deepmriprep, termed supervised SYMNet (sSYMNet), addresses these issues and enables prediction of smooth deformation fields on par with CAT12. Although median metrics slightly favour CAT12, visual inspection suggests that the remaining performance gap is minor, even in challenging cases.

VBM analysis results using deepmriprep and CAT12 show high similarity, with respective t-maps typically showing strong correlation coefficients of over 0.8. This correlation remains robust for the investigated psychometric variables, years of education, HC vs. MDD, and IQ, despite their smaller effect sizes compared to biological variables age, sex, and BMI. Comparable maximal t-scores across all investigated variables further support the equivalence of outcomes between deepmriprep and CAT12. The largest differences in maximal t-scores and the lowest t-map correlation occurs for BMI and is mainly caused by a cluster in the outer cerebellum identified only by CAT12.

The high processing speed of deepmriprep addresses a critical need in VBM analysis, where MRI studies with over 1,000 images are often necessary to accurately measure effect sizes \cite{marek2022}. For instance, preprocessing the largest currently existing study containing more than 100,000 MR images \cite{bycroft_uk_2018} using CAT12 preprocessing would require over 6 months of computation time on our high-end system\footnote{CPU: AMD Ryzen 9 5950X, RAM: 128GB, GPU: NVIDIA GeForce RTX 3090}. In contrast, deepmriprep can complete the same task in just five days on the same hardware. By alleviating the processing bottleneck in VBM preprocessing, deepmriprep accelerates scientific progress, enabling laboratories without extensive computational resources to efficiently handle large-scale studies.

Although deepmriprep’s high processing speed and user-friendly interface are its main advantages, its underlying software design may hold even greater significance for future development (see \url{https://github.com/wwu-mmll/deepmriprep}). The software is organized into small, modular components that consist of less than 1,000 lines of code. This streamlined design enhances long-term maintainability and reduces the likelihood of potentially far-reaching bugs \cite{afni, eklund_cluster_2016}. Most importantly, deepmriprep’s straightforward software architecture lowers the barrier for researchers in VBM and other neuroimaging domains, making it easier to understand, adapt and reuse the code for various neuroimaging pipelines. We anticipate that the broader adoption of deepmriprep into other neuroimaging pipelines will advance the underlying methods, thereby fostering progress across the broader neuroscience community.

\section*{Declarations}
\subsection*{Conflict of Interest}
All authors declare that they have no conflicts of interest.

\subsection*{Funding}
This work was funded by the German Research Foundation (DFG grants HA7070/2-2, HA7070/3, HA7070/4 to TH) and the Interdisciplinary Center for Clinical Research (IZKF) of the medical faculty of Münster (grants Dan3/012/17 to UD and MzH 3/020/20 to TH). This work was supported in part by the SFB/TRR 393 consortium from the German Research Foundation (DFG). The BiDirect Study is supported by grants of the German Ministry of Research and Education (BMBF) to the University of Muenster (01ER0816 and 01ER1506).

\section*{Supplements}
\setcounter{figure}{0}
\renewcommand{\thefigure}{S\arabic{figure}}
\setcounter{table}{0}
\renewcommand{\thetable}{S\arabic{table}}
\setcounter{section}{0}
\renewcommand{\thesection}{S\arabic{section}}

\begin{table}[H]
\caption{Magnetic Resonance Imaging (MRI) scanner models occurring in the compilations of datasets OpenNeuro-Total and OpenNeuro-HD.}
\label{tab:scanner}
\centering
\begin{tabular}{l|rr|rr}
{} &  \multicolumn{2}{c}{\textbf{OpenNeuro-Total}} &  \multicolumn{2}{c}{\textbf{OpenNeuro-HD}} \\
\textbf{MRI scanner model} & No. of datasets & No. of subjects & No. of datasets & No. of subjects \\
\midrule
General Electric Discovery &          15 &                          360 &                         8 & 40 \\
General Electric Excite    &           1 &                           15 &                         0 & 0 \\
General Electric Signa     &           3 &                           77 &                         0 & 0 \\
General Electric Unknown   &           1 &                           38 &                         0 & 0 \\
Philips Achieva            &          18 &                         1840 &                        10 & 50 \\
Philips Achieva X          &           4 &                          149 &                         4 & 20 \\
Philips Ingenia            &           5 &                          237 &                         4 & 20 \\
Philips Intera             &           5 &                           97 &                         1 & 5 \\
Siemens Allegra            &          14 &                          264 &                         0 & 0 \\
Siemens Avanto             &           4 &                          175 &                         4 & 20 \\
Siemens Biograph           &           3 &                          105 &                         1 & 5 \\
Siemens Magnetom           &           5 &                           95 &                         0 & 0 \\
Siemens Prisma             &          43 &                         1531 &                        33 & 165 \\
Siemens Skyra              &          18 &                          944 &                        18 & 90 \\
Siemens Trio               &          55 &                         1964 &                        47 & 235 \\
Siemens Unknown            &           1 &                           49 &                         1 & 5 \\
Siemens Verio              &           8 &                          166 &                         4 & 20 \\
Siemens Vida               &           1 &                           50 &                         1 & 5 \\
Unknown                    &           4 &                          122 &                         1 & 5 \\
\end{tabular}
\end{table}

\begin{figure}[H]
	\centering
	\includegraphics[width=\textwidth]{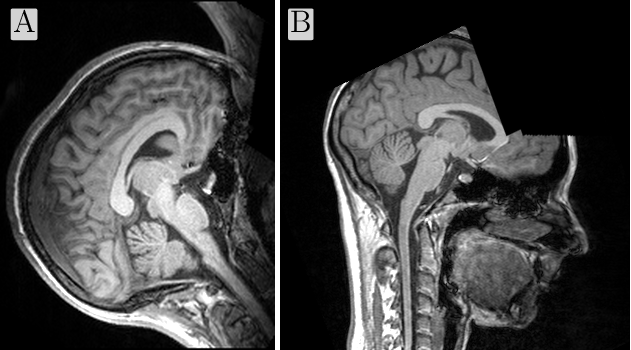}
	\caption{Image dropouts caused by erroneous orientation (A), and improper masking (B).}
	\label{fig:dropouts}
\end{figure}

\begin{figure}[H]
	\centering
	\includegraphics[width=0.82\textwidth]{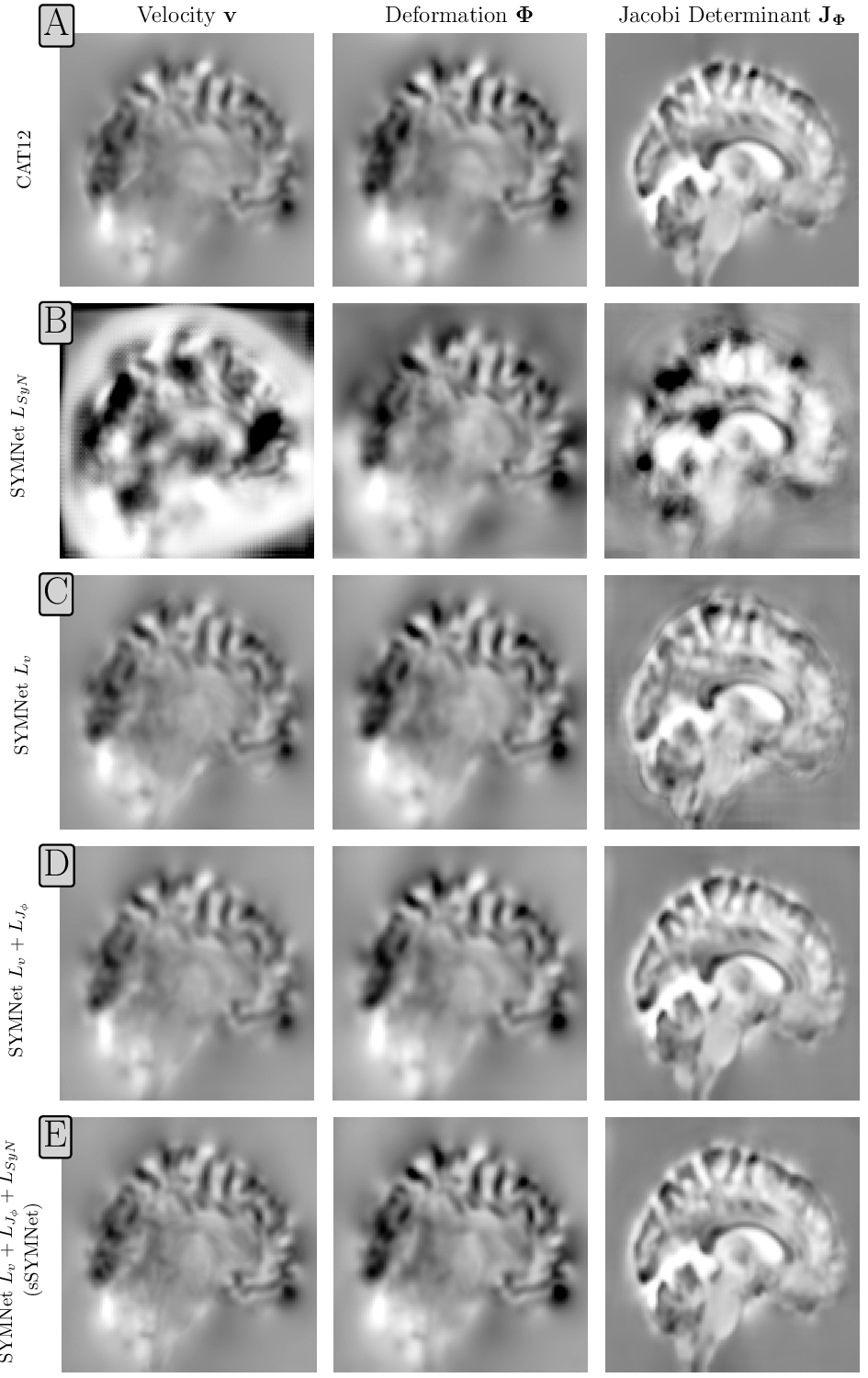}
	\caption{Sagittal displacements in the velocity field, deformation field, and Jacobi determinant of one example image produced by CAT12 (A) and four variants of SYMNet trained with different loss functions (B-E).}
	\label{fig:ablations}
\end{figure}

\begin{figure}[H]
	\centering
	\includegraphics[width=\textwidth]{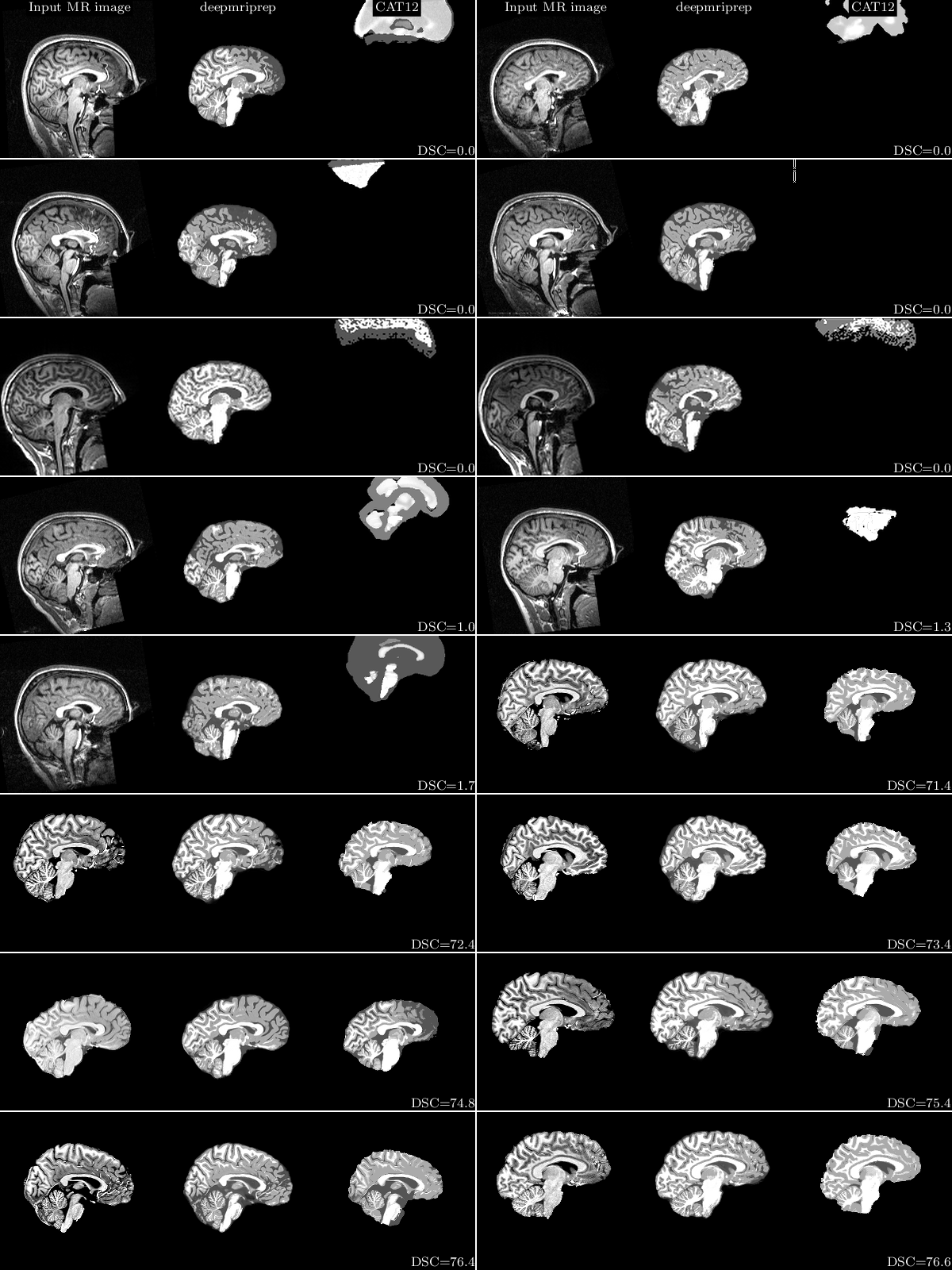}
	\caption{16 out of 8,279 MR images from OpenNeuro-Total which resulted in the tissue maps with the largest disagreement - i.e., lowest Dice score with respect to gray and white matter - between deepmriprep and CAT12.}
	\label{fig:p0totalsupp}
\end{figure}

\begin{figure}[H]
	\centering
	\includegraphics[width=0.84\textwidth]{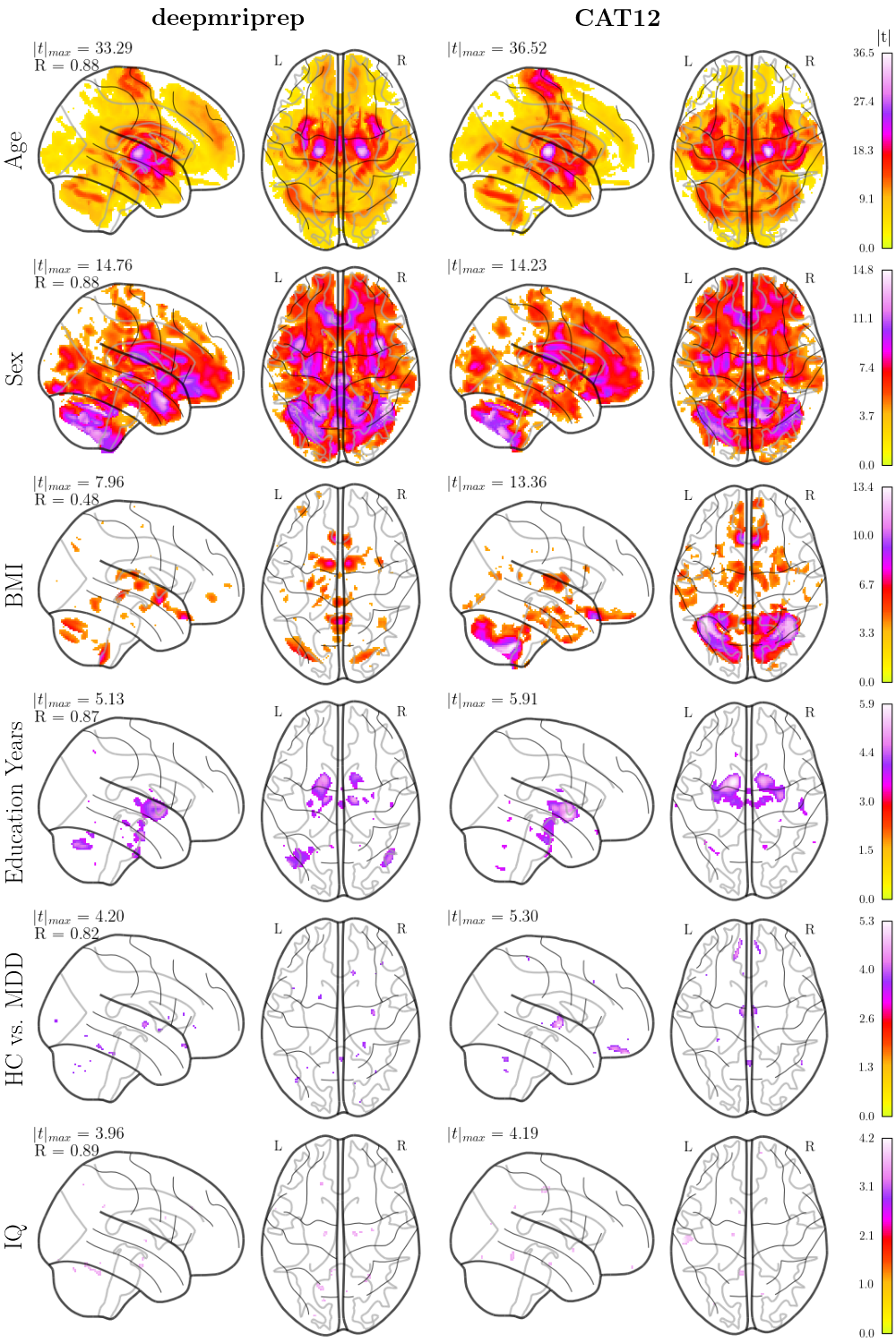}
	\caption{Absolute t-scores of GLM analysis between white matter volume and age, sex, Body mass index (BMI), years of education, HC vs. MDD (healthy control vs. major depressive disorder) and intelligence quotient (IQ) based on deepmriprep- (left) and CAT12-preprocessing (right) thresholded at $p<0.001$. The respective maximum values and correlation coefficients between deepmriprep and CAT12 are based on unthresholded absolute t-scores.}
	\label{fig:tscoreswm}
\end{figure}

\subsection{Gray Matter Masking} \label{sec:nogm}
Based on SPM, CAT12 outputs tissue segmentation maps (file-prefix “p0”) with continuous values ranging from 0 to 3. The values 0, 1, 2 and 3 encode the segmentation classes background, cerebrospinal fluid (CSF), gray matter (GM), and white matter (WM). Intermediate values, like e.g. 2.4, can thereby easily be mapped to the respective voxel containing 40\% gray matter and 60\% white matter. Applying this mapping to all voxels results in the tissue probability maps for GM, WM, and CSF (file-prefixes “p1”, “p2”, and “p3”). By comparing these probability maps with the respective segmentation map, we found voxels positioned on the edge of the ventricles and the brain stem which did not follow this mapping (see Figure S5). The GM probability in these voxels is set to zero, and the WM and CSF probabilities are each increased by half of the original GM probability to ensure a tissue probability sum of 100\%.

\begin{figure}[H]
	\centering
	\includegraphics[width=\textwidth]{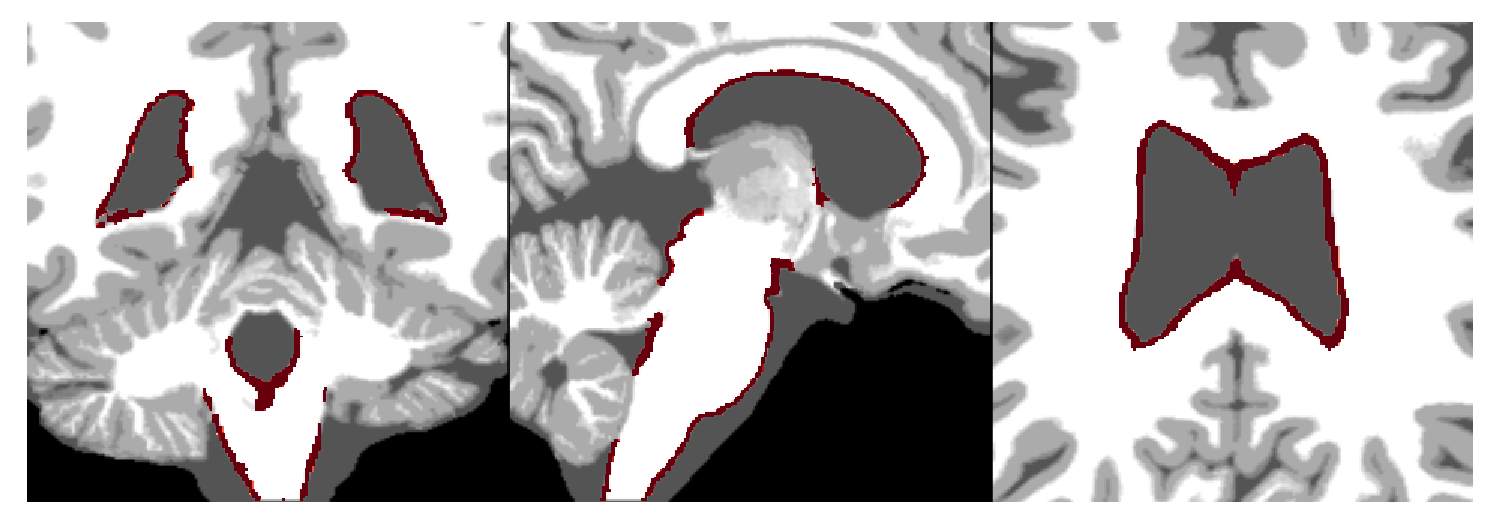}
	\caption{Coronal (left), sagittal (middle), and axial slice (right) of an example tissue segmentation map, with all voxels subject to gray matter (GM) masking highlighted in red.}
	\label{fig:gmline}
\end{figure}

To make deepmriprep conformant with this GM masking, we employed the UNet architecture already used for tissue segmentation (see Section 2.2.1) to predict the corresponding voxels that would be masked in CAT12. As model input, a 224x288x256 voxel region of the tissue segmentation map with a resolution of 0.5mm is used. We follow a patchwise approach similar to the tissue segmentation model with two instead of 27 static patch positions (see Figure S6). The two patches each cover 128x288x256 voxels and are placed symmetrically on the left and right hemispheres. We flip all right hemisphere patches along the sagittal axis during model training so that the resulting model can be used to predict the GM mask in both hemispheres.

\begin{figure}[H]
	\centering
	\includegraphics[width=\textwidth]{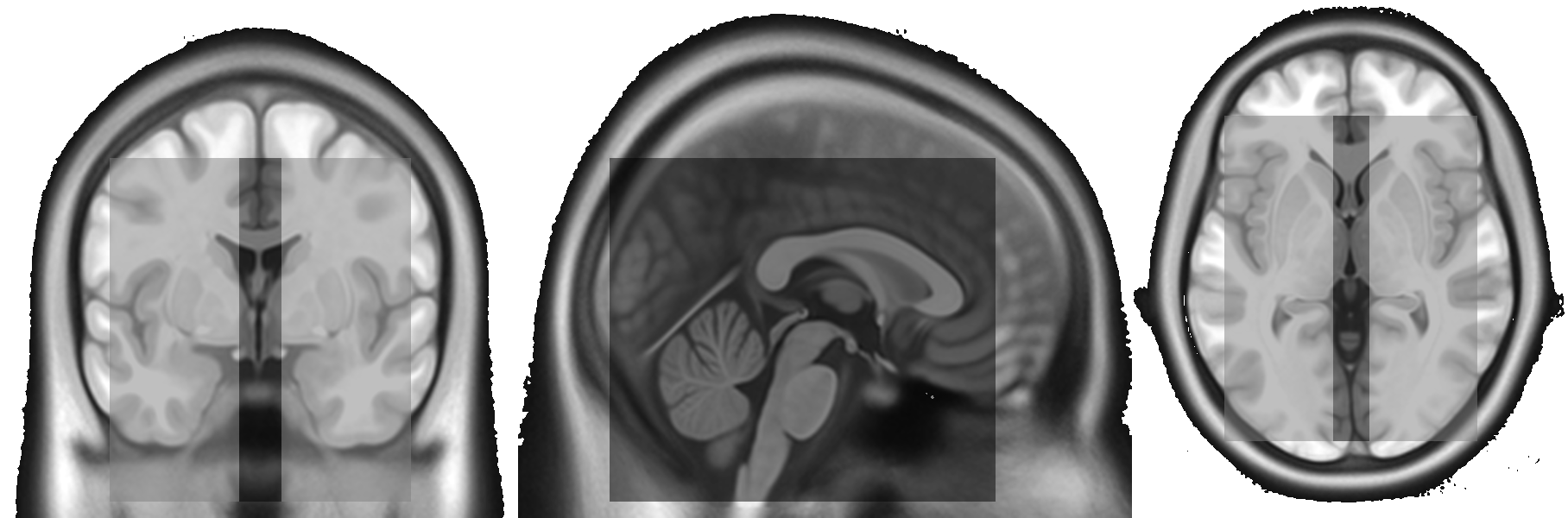}
	\caption{Coronal (left), sagittal (middle), and axial slice (right) of the two 128x228x256 voxel patches used for gray matter masking. For reference, the T1 template of CAT12, upsampled to the utilized resolution of 0.5mm, is shown in the background}
	\label{fig:nogmpatches}
\end{figure}

\begin{figure}[H]
	\centering
	\includegraphics[width=\textwidth]{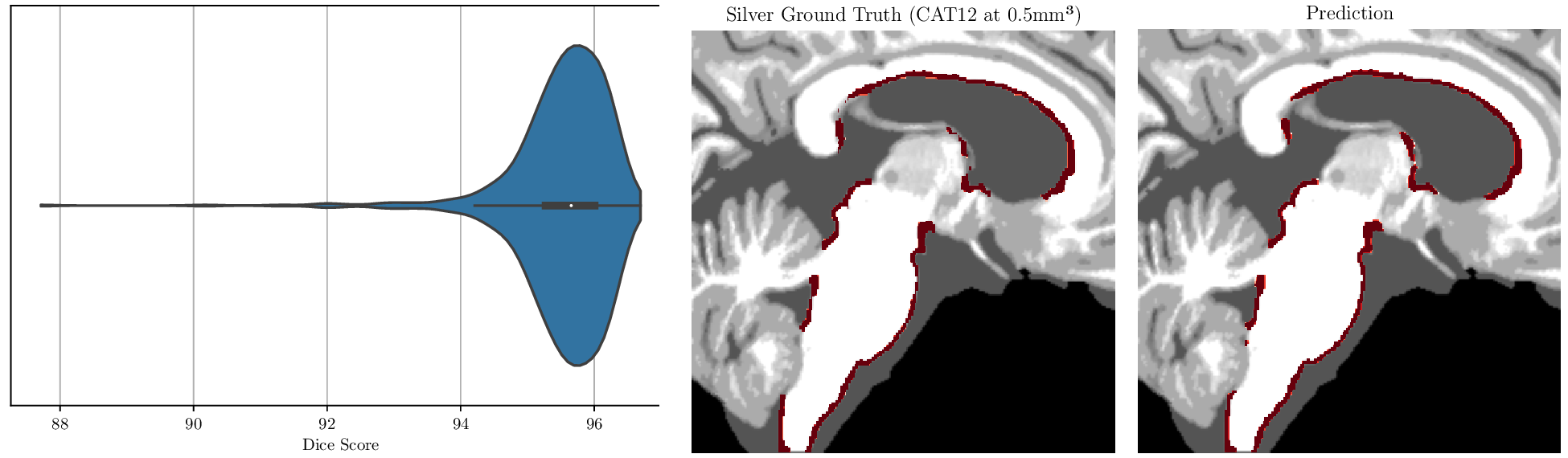}
	\caption{Left: Distribution of Dice scores of the predicted gray matter (GM) masks across all validation images obtained during 5-fold cross-dataset validation. Right: Sagittal slice of the predicted GM mask (red), which resulted in the lowest Dice score compared to the ground truth GM mask (red). The model input, i.e. the respective tissue segmentation map, is shown in the background.}
	\label{fig:nogm}
\end{figure}

\end{document}